\documentclass[acmtog, nonacm]{acmart}

\AtBeginDocument{%
  }

\setcopyright{cc}
\setcctype{by-sa}
\acmJournal{TOG}
\acmYear{2025} \acmVolume{44} \acmNumber{6} \acmArticle{212} \acmMonth{12} \acmPrice{}\acmDOI{10.1145/3763294}

\acmSubmissionID{1287}
\usepackage{graphicx}
\usepackage{caption}
\usepackage{float}
\usepackage{subcaption}

\usepackage[linesnumbered,ruled,vlined]{algorithm2e}
\usepackage{algpseudocode}  
\usepackage{soul}
\newcommand{\Sect}[1]{Sec.~\ref{#1}}

\newcommand{\Fig}[1]{Fig.~\ref{#1}}
\newcommand{\Tbl}[1]{Table~\ref{#1}}
\newcommand{\Equ}[1]{Eq.~\ref{#1}}
\newcommand{\Apx}[1]{Supplementary Material~\ref{#1}}
\newcommand{\Alg}[1]{Algo.~\ref{#1}}

\DeclareRobustCommand{\fixedme}[1]{#1
}

\newcommand{\fixme}[1]{}

\newcommand{\no}[1]{}

\renewcommand{\paragraph}[1]{\textbf{#1} }

\def\cS{{\texttt{c}_\texttt{S}}}
\def\cT{{\texttt{c}_\texttt{T}}}

\def\mP{{\mathcal{P}}}

\def\tT{{\mathtt{T}}}
\def\tS{{\mathtt{S}}}
\def\d65{{D65}}
\def\bd65{{\mathbf{d_{65}}}}
\def\upvp{{$u'v'$}}

\makeatletter
\newcommand\thefontsize{The current font size is: \f@size pt}
\makeatother

\DeclareMathOperator*{\argmax}{arg\,max}
\DeclareMathOperator*{\argmin}{arg\,min}

\begin{document}

\title{Modeling and Exploiting the Time Course of Chromatic Adaptation for Display Power Optimizations in Virtual Reality}




\begin{abstract}

We introduce a gaze-tracking--free method to reduce OLED display power consumption in VR with minimal perceptual impact.
This technique exploits the time course of chromatic adaptation, the human visual system’s ability to maintain stable color perception under changing illumination.
To that end, we propose a novel psychophysical paradigm that models how human adaptation state changes with the scene illuminant.
We exploit this model to compute an optimal illuminant shift trajectory, controlling the rate and extent of illumination change, to reduce display power under a given perceptual loss budget.
Our technique significantly improves the perceptual quality over prior work that applies illumination shifts instantaneously.
Our technique can also be combined with prior work on luminance dimming to reduce display power by 31\% with no statistical loss of perceptual quality.

\end{abstract}

\begin{teaserfigure}
    \centering
    \includegraphics[width=\linewidth]{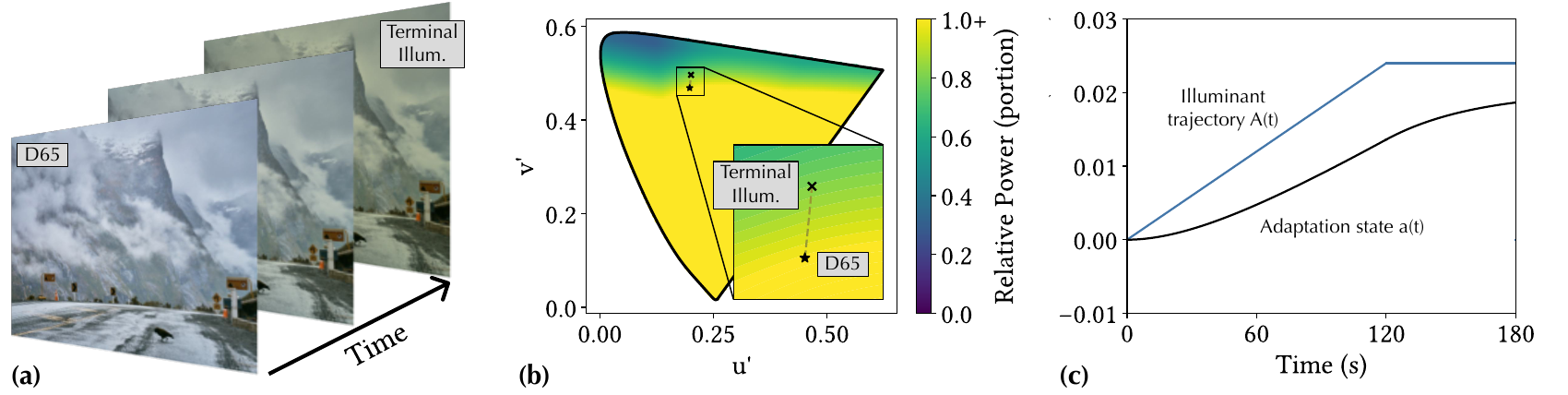}
    \captionsetup{justification=raggedright, singlelinecheck=false} 
    \caption{\textit{Illustration of our gaze-tracking--free VR display power saving method exploiting the time course of human chromatic adaptation.} (a): Our algorithm gradually shifts the illuminant in the virtual scene from the CIE Standard Illuminant \d65 (average daylight) to a green-ish tint.
    This is done by carefully modulating the trajectory and the rate of the illuminant traversal to maximize display power saving while minimizing perceptual impact.
    (b): The plot shows the trajectory of the illuminant traversal applied in (a) in the CIE \upvp chromaticity space.
    The colors in the diagram represent the estimated display power consumptions under different illuminants; the power is normalized relative to \d65, and clipped at 1.
    All the illuminants that yield power savings relative to \d65 are colored green; the illumination shift in (a) leads to about 31\% power saving.
    (c): To bound perceptual loss during the illuminant shift, we propose a novel psychophysical paradigm to model human adaptation state under illumination changes.
    The model predicts the user adaptation state $a(t)$, (i.e., one's endogenous notion of neutral illuminant) given the actual illuminant $A(t)$ in the scene.
    We bound the difference between the internal adaptation state and the actual illuminant to ensure perceptual quality during illuminant shifts.
    Credit for the original image used in Fig. 1a goes to Pedro Szekely~\cite{kea_borb}.
    }{}
    \label{fig:teaser}
\end{teaserfigure}

\author{Ethan Chen}
\email{echen48@ur.rochester.edu}
\orcid{0009-0008-1250-769X}
\affiliation{
    \institution{University of Rochester, Meta}
    \city{Rochester}
    \state{New York}
    \country{United States of America}
}

\author{Sushant Kondguli}
\email{sushantkondguli@meta.com}
\orcid{0000-0002-7295-4626}
\affiliation{
    \institution{Meta}
   \city{Sunnyvale}
    \state{California}
    \country{United States of America}
}

\author{Carl Marshall}
\email{csmarshall@meta.com}
\orcid{0009-0001-7288-5341}
\affiliation{
    \institution{Meta}
    \city{Redmond}
    \state{Washington}
    \country{United States of America}
}

\author{Yuhao Zhu}
\email{yzhu@rochester.edu}
\orcid{0000-0002-2802-0578}
\affiliation{
    \institution{University of Rochester}
    \city{Rochester}
    \state{New York}
    \country{United States of America}
}


\maketitle

\section{Introduction}
Virtual reality (VR) devices are severely power constrained.
As the need for higher frame rate and higher resolution keeps increasing, the display contributes to an increasingly significant portion of the total power consumption~\citep{duinkharjav2022color, leng2019energy}, and is a prime candidate for power optimization.



\fixedme{As long been established, OLED display power consumption is color-dependent~\citep{dong2009power, dash2021much, chen2024pea, duinkharjav2022color}.
\no{This is because different subpixels (display primary lights) in OLED displays consume different amounts of power.
As different pixel colors require different combinations of display primary lights to reproduce, they can consume different display power (even when controlling for luminance).}
This paper presents a new method to exploit the color dependency of OLED display power consumption.
Our technique relies on chromatic adaptation, the process by which our visual system scales it sensitivity to maintain a stable color perception under variable lighting conditions~\cite{color_appearance_models} (\Sect{sec:bck}).
For example, when moving from warm, incandescent lighting indoors to cooler sunlight outdoors, our visual system adapts such that neutral objects, e.g., a white sheet of paper, always appear white-ish, even though the ``objective'' color of the paper in the colorimetric sense has changed.
\no{Chromatic adaptation occurs because the spectral sensitivity of our cone photoreceptors adjust according to the illuminant as if the illuminant is ``discounted''.}
}

We exploit this phenomenon in virtual scenes.
Given a reference scene rendered under an illuminant $\texttt{S}$, our idea is to render the same scene with a different illuminant $\texttt{T}$;
the new rendering accounts for the photoreceptor sensitivity change from $\texttt{S}$ to $\texttt{T}$ and, accordingly, shifts all the pixel colors such that the new rendering 1) appears to be perceptually equivalent to the reference rendering and 2) reduces the display power consumption.
The pixel color shift can be implemented as a real-time shader with little computational cost.

While conceptually simple, leveraging chromatic adaptation is challenging, because adaptation is not an instantaneous phenomenon --- the human visual system can take several minutes to fully adapt to a change in illuminants~\cite{gupta_5_mins, fairchild1995time}.
The gradual nature of chromatic adaptation means that instant shifts in the illumination of VR environments are noticeable, as previously demonstrated~\citep{chen2024pea}.

Our goal, thus, is to devise a principled strategy to gradually adapt the scene illuminant such that the perceptual impact is minimized, while power savings are maximized. 
This entails two tasks.
First, we must identify ``lucrative'' illuminants that, if shifted to, would lead to significant power savings.
We identify such illuminants using the chromatic adaptation theory and display power modeling (\Sect{sec:illu}).

\fixedme{Second, we must determine \textit{how} the current illuminant should be shifted.
\no{There exists a trade-off between power savings and perceptual fidelity.
An illuminant that is extremely far away from normal daylight (e.g., CIE Standard Illuminant \d65) may be power-efficient, but cannot be completely adapted to by users. 
Similarly, a faster shift toward a power-efficient illuminant yields power savings faster, but might not give users sufficient time to adapt.}
To that end, we propose a novel psychophysical paradigm that can query the rate and bounds of a user's adaptation state in real-time.
Using observations from the psychophysical study, we build a statistical inference model using Maximum Likelihood Estimation that estimates how a user's adaptation state changes over time, given the trajectory of the illuminant change (\Sect{sec:exp}).
}

Using the statistical model, we formulate a power optimization problem to calculate the illuminant shift trajectory that minimizes power consumption under a given perceptual loss budget.
The optimal trajectory is then applied online during rendering time; this is implemented as a real-time fragment shader using little compute budget on a mobile VR GPU (\Sect{sec:power}).

Through extensive subjective studies on a Meta Quest Pro, we demonstrate that our algorithm leads to significantly better perceptual quality compared to a previous chromatic-adaptation--based power saving method under the same power budget (\Sect{sec:validation}).
When combined with uniform luminance dimming, our method achieves 31\% power saving with no statistical degradation in perceptual quality.
\fixedme{We also discuss the limitations of our method (e.g., when the adaptation state has to be reset) and how it could be extended to accommodate more diverse VR content (\Sect{sec:disc}).}

%
In summary, this paper makes the following contributions:
\vspace{-2pt}
\begin{itemize}
    \item We analyze the power-saving opportunities under chromatic adaptation by mining a natural-scene dataset.
    \item We propose a novel psychophysical paradigm and statistical inference method to probe and model human adaptation state under dynamic (time-varying) illumination.
    \item Leveraging the computational model, we propose a new display-power saving algorithm that dynamically shifts the scene illuminant to minimize power consumption under a perceptual degradation bound.
    \item Through subjective validation studies, we show that our algorithm saves about 31\% display power with no statistical loss in perceptual quality.
\end{itemize}



\section{Preliminaries on Chromatic Adaptation}
\label{sec:bck}

Throughout the day, we encounter a variety of lighting conditions, from incandescent bulbs to natural sunlight, yet our color perception remains relatively stable.
This constancy arises partially because our visual system adapts to differing lighting conditions through a process known as chromatic adaptation.
We discuss the key concepts necessary for the rest of the paper here, and leave the mathematical details to \Apx{sec:cat}.


\citet{kries1902chromatic, kries1905influence} hypothesized that (in modern interpretations) we adapt by scaling the spectral sensitivities of individual classes of cone photoreceptors under different illuminants\footnote{This is a phenomenological model in that sensitivity scaling in other retinal cells are also almost certainly involved~\citep{webster2011adaptation}.}.
The sensitivities scale such that the cone responses of a neutral point remain constant under different illuminants as if the color of the illuminant is ``discounted.'' 
A neutral point reflects lights uniformly across its wavelength, so its color is the color of the illuminant itself.
Since common illuminants (e.g., daylights) appear white-ish, the color of the illuminant or the color of a neutral point is also called the ``\textbf{white point}'' or ``reference white'' of the scene.

An important concept in chromatic adaptation is the user's \textbf{adaptation state}, or the ``internal white point'', which refers to the color that the user endogenously regards as white.
When a user is exposed to a natural illuminant (e.g., daylight) $\tS$ for a sufficiently long time, the user has fully adapted to the illuminant, at which point the user's adaptation state \textit{is} $\tS$.

Critically, the adaptation state is not always the same as the illuminant.
First, adaptation takes time (partially because it takes several minutes for the cone sensitivities to settle when exposed to a new illuminant)~\citep{fairchild1995time, gupta_5_mins}.
Second, while humans adapt to different daylights reasonably well (presumably driven by evolution~\cite{daylight_better, radonjic_daylight}), we do not adapt fully to overly colored illuminants: imagine entering a room illuminated by neon red; under no amount of exposure, however long, will we perceive the walls as white.
A main focus of this paper is to model the time course of the adaptation state, even when the illuminant itself is changing.

Adaptation state must be considered during rendering.
If an image is initially presented to a user adapted to illuminant $\tS$ but then later presented to the same user, now adapted to a different illuminant $\tT$, the colors in the image will appear differently as they do under $\tS$, as cone sensitivity scaling depends on the adaptation state.

To compensate for the change of adaptation state, we must transform a color $\cS$ in the original image rendered under $\tS$ to $\cT$ such that $\cT$, viewed under $\tT$, appears to be the same color as $\cS$, viewed under $\tS$.
This conversion is done via a linear Chromatic Adaptation Transform (CAT) function, whose details are described in \Apx{sec:cat} and will be denoted as $f_{\tS\rightarrow\tT}(\cdot)$ in the paper:
\begin{align}
\begin{aligned}
\label{eq:cat_srgb}
    \cT = &f_{\tS\rightarrow\tT}(\cS).
\end{aligned}
\end{align}

\fixedme{A} visual example of how colors are transformed is graphed in the CIE \upvp chromaticity space in \Fig{fig:color_shifts_adapt}, where we assume the source illuminant $\tS$ is the CIE Standard Illuminant \d65 and the target illuminant $\tT$ is a yellow-ish color.
\d65 approximates the average daylight, and is the white point in almost all common color spaces.
Overlaid in the graph are the sRGB color gamut and the Display P3 color gamut, both of which use \d65 as the white point.
We uniformly sample the colors in the sRGB gamut. The arrows associated with each sRGB color represent how the color shifts when chromatically adapted from \d65 to $\tT$.

A yellow-ish $\tT$ is not a random choice.
Prior measurements on an OLED display shows that blue-ish colors consume more power on that display~\citep{chen2024pea, duinkharjav2022color}.
Our results here show that, for that particular display, shifting the illuminant to a yellow-ish color would shift all colors away from blue, leading to power savings.
We will later quantify the power saving under different target illuminants (\Sect{sec:illu}).

\begin{table*}[!t]
    \caption{A table of symbols used in the paper and their definitions.}.
    \centering
    \begin{tabular}{c|c}
        \toprule[0.10em]
         \textbf{Symbol} & \textbf{Definition} \\
         \midrule[0.05em]
         $f_{\texttt{S}\rightarrow \texttt{T}}(\texttt{c})$ & CAT function transforming a linear sRGB color $\texttt{c}$ from illuminant \texttt{S} to a new illumination $\texttt{T}$.  \\ 
         $p(\texttt{c})$ & Display power consumption of displaying color $\texttt{c}$.\\ 
         $\mathbf{p}_{disp}$ & Vector of unit power of each color channel.\\ 
         $p_{static}$ & Static power of a display.\\ 
         $\mathcal{P}(\texttt{T})$ & Average display power consumption under illuminant $\texttt{T}$.\\ 
         $A(t)$ & The scene illumination as a function of time $t$.\\ 
         $a(t)$ & A user's adaptation state as a function of time $t$.\\ 
         $a'(t)$ & The rate of change in the user's adaptation state over time $t$. \\
         $k_1$ & Coefficient of adaptation rate.\\ 
         $k_2$ & Coefficient of adaptation completeness $\in[0, 1]$.\\ 
         $F, L$ & Two stimuli used in 2AFC tests in \Sect{sec:exp}. $F$ (farther) is ahead of $L$ (lagging) in the direction of illuminant traversal.\\ 
         $m$ & Midpoint between $F$ and $L$.\\ 
         $G(m-a; \theta)$ & A logistic psychometric function showing the proportion of picking $L$ given $m-a$; it is parameterized by $\theta = \{k, x_0\}$.\\ 
         $v$ & Velocity of illuminant traversal defined as \upvp distance/second.\\ 
         $\phi$ & Direction of illuminant traversal defined in radians (horizontal direction pointing to the right is defined as $\phi=0$).\\ 
         $t_1$, $t_2$ & Duration of the second and third measurement phase of the pilot study (\Sect{sec:exp:proc}).\\ 
         $D$ & Distance of illuminant traversal in the first measurement phase of the pilot study (\Sect{sec:exp:proc}).\\ 
         $\bd65$ & Chromaticity of \d65 in the \upvp space ($\in \mathbb{R}^2$).\\ 
         $\Delta T$ & Permissible difference between $A(t)$ and $a(t)$ in the power optimization (\Sect{sec:power}).\\ 
         \bottomrule[0.10em]
    \end{tabular}
    \label{tab:symbol}
\end{table*}


\section{Related Work}
\label{sec:related}

\subsection{Time Course of Chromatic Adaptation}
\label{sec:related:ca}

Chromatic adaptation is not instantaneous ~\citep{color_appearance_models}.
The adaptation state of a human shifts slowly in response to changes in illumination.
Several studies have measured the time course of chromatic adaptation under a stable illuminant after an instantaneous change\fixedme{~\citep{rinner_time_2000, fairchild1995time, hunt_effects_1950, shevell_time_2001, jameson_receptoral_1979, coia2024measurements, belmore2011very}}. 
Our focus, however, is to model the time course of adaptation under dynamic (time-varying) illumination.

Only two studies to date have focused on gradual illuminant changes \citep{temporal_daylight, time_course_chromatic}. 
These studies, however, have limitations in their methodologies that we address.

\citet{time_course_chromatic} studies chromatic adaptation under smooth transitions in illumination along a blue-yellow trajectory.
At different times during the transition, a participant is presented with a colored stimuli, and is asked to classify the color as either bluish or yellowish.
To estimate the adaptation state, the authors assume that if a queried color is achromatic (matching the adaptation state), the participants would have an even chance of classifying the color as either yellowish or bluish.
Their task design has two challenges.
First, the task relies on color naming, which is subjective and culturally dependent~\cite{kay2003resolving, gibson2017color}.
Second, as we show in \Sect{sec:exp:cal}, the assumption that the achromatic point corresponds to a 50/50 chance of picking yellow or blue is questionable.

\citet{temporal_daylight} also measures chromatic adaptation under dynamic lighting.
The illuminant is shifted between different colors along the daylight curve over a 10 second span, by the end of which participants are asked if they noticed any change in illumination.
The paper does not aim to model the user adaptation state as a function of changes in lighting conditions, which is a main focus of our work.
Due to the hard 10-second limit, their psychophysical paradigm is limited in its ability to probe the bounds of illuminants that humans can adapt to.
We derive such bounds, which is both of scientific value, and directly impacts our power savings technique.

\subsection{Display Power Optimization}
\label{sec:related:power}

Reducing display power has long been a subject of research; techniques include reducing luminance~\citep{dash2021much, shye2009into, yan2018exploring}, better hardware design~\citep{miller2006p, shin2013dynamic}, color modulation~\citep{dong2009power, dong2011chameleon}, and multi-primary displays~\citep{boroson2009oled, miller2007color}.
Recently, there is been an interest in VR display power optimizations~\citep{chen2024pea, duinkharjav2022color}.

\no{\citet{duinkharjav2022color} exploited the human visual system's reduced color discrimination in peripheral vision to shift rendered colors imperceptibly while reducing display power.
Their work builds a power model for OLED displays, which we use in this paper.
A key advantage of our technique compared to ~\citep{duinkharjav2022color} is that we do not rely on gaze tracking, which could be power hungry~\citep{chen2024pea, feng2024blisscam}.
However, color foveation and chromatic adaptation are not mutually exclusive.
To systematically exploit the two simultaneously, an interesting future work is to derive color discrimination thresholds under different illuminants and adaptation states.
}

PEA-PODs~\citep{chen2024pea} evaluates the perceptual impact of many power-saving algorithms. 
One of the algorithms tested is ``whitepoint shift'', which, as we discuss in \Sect{sec:validation:comparison}, applies the illuminant shift instantaneously.
This results in visible artifacts, as their implementation neglects the gradual nature of chromatic adaptation, which is what this paper aims to address.

Additionally, \citet{chen2024pea} does not construct or use any models of chromatic adaptation to guide their white-point shift algorithm.
Instead, they re-use the color discrimination model from \citet{duinkharjav2022color} to decide on an optimal illuminant by setting the eccentricity to $0\deg$.
In contrast, this paper proposes a novel psychophysical paradigm to model the human adaptation state and use the model to derive an optimal illumination shift.

\subsection{Models of Perceptual Quality}
\label{sec:related:perceptual}

Many previous works have characterized and modeled human visual perception under various low-level stimulus characteristics such as spatial/temporal frequency, \fixedme{stimuli area}, luminance, eccentricity, color, \fixedme{and background conditions}~
\fixedme{\citep{ashraf2024castlecsf, mantiuk2022stelacsf, cai2024elatcsf, krajancich2021perceptual, tursun2022perceptual}}.
These models could serve as the basis of subjective perceptual metrics~\citep{mantiuk2021fovvideovdp, colorvideovdp}.
Many perceptual metrics consider mid to high-level 
processing in the visual system~\citep{walton2021beyond, fu2023dreamsim}.

Some of the quality metrics have been extended to work on Augmented Reality data~\citep{ardavid}.
To our best knowledge, none accounts for (shifts in) the adaptation state.
\citet{colorvideovdp} explicitly assumes that the user is adapted to \d65 lighting conditions.
It would be an interesting future work to integrate the time course of chromatic adaptation into subjective metrics.

Color Appearance Models~\citep{color_appearance_models} do account for the scene illuminants but their focus is on modeling steady-state adaptation under stable illuminants.
Our focus, however, is to model the dynamics of adaptation under dynamic illumination.

\section{Identifying Power-Saving Illuminants}
\label{sec:illu}

The goal of this section is to identify ``lucrative'' illuminants under which one could obtain significant power savings.
This section is not concerned with \textit{how} the illuminant should be adjusted (to minimize perceptual impact), which is the focus of \Sect{sec:exp}.

\subsection{Methodology}
\label{sec:illu:method}

In order to identify power-saving illuminants, we must build a model that relates the display power consumption to the target illuminant \texttt{T} we want to change to.
Fundamentally, however, the power consumption of emissive displays like an OLED is dictated by the pixel colors.
We relate the power consumption to the illuminant of a virtually rendered scene in three steps, as shown in \Equ{eq:power_illu}.
\begin{subequations}
\label{eq:power_illu}
\begin{align}
    \cT &= f_{\d65\rightarrow \texttt{T}}(\texttt{c}), \label{eq:power_illu_b} \\
    p(\cT) &= \mathbf{p}_{disp}^T \times \texttt{c}_\texttt{T} + p_{static}, \label{eq:power_illu_c} \\
    \mP(\texttt{T}) &= \sum^N_\texttt{c}p(\cT)\cdot W_\texttt{c}. \label{eq:power_illu_a}
\end{align}
\end{subequations}
First, we calculate how each linear sRGB color $\texttt{c}\in \mathbb{R}^3$ in the original scene is adapted from \d65 illuminant to a new illuminant $\texttt{T} \in \mathbb{R}^3$, assuming that the adaptation to $\texttt{T}$ is complete.
We choose \d65 as the source illuminant because it is the white of all commonly used color spaces supported by modern displays, and it is assumed that users are adapted to \d65 when viewing content exclusively from emissive displays (without ambient lights), which is the case in VR~\citep{color_appearance_models}.
The adaptation is carried out by the CAT function $f_{\d65\rightarrow \texttt{T}}(\cdot)$ in \Equ{eq:power_illu_b} (see \Sect{sec:bck}).
Let's denote the adapted color as $\cT$.

Second, we then estimate the power consumption of displaying $\cT$.
The power consumption of displaying a particular color on OLEDs is established in the literature~\citep{dong2009power, dash2021much, tsujimura2017oled} as a linear combination of the RGB channel values in the color plus the constant static power $p_{static}$ (e.g., consumed by peripheral circuitry~\citep{huang2020mini}).
This is shown in \Equ{eq:power_illu_c}, where $\mathbf{p}_{disp} \in \mathbb{R}^3$ represents the vector of the unit power of each color channel, and is usually regressed from actual display power measurement.
In our study, we use the parameters measured and reported on a Wisecoco 3.81 inch OLED Display used in prior work~\citep{duinkharjav2022color, chen2024pea}.

Finally, since the pixel colors in an image to be displayed depend on the specific scene being rendered, the power saving is necessarily scene specific.
For the results to be generally applicable, we instead estimate an average power saving given natural scene statistics.

\begin{figure*}[t] 
    \centering
    \begin{subfigure}{\linewidth}
        \centering
        \includegraphics[width=\linewidth]{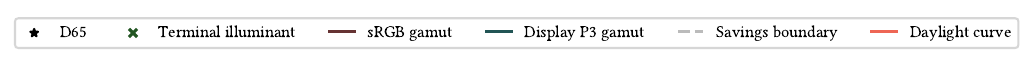}
    \end{subfigure}\\
    \subfloat[Color shifts under chromatic adaptation.]
    {
    \label{fig:color_shifts_adapt}
    \includegraphics[height=1.75in]{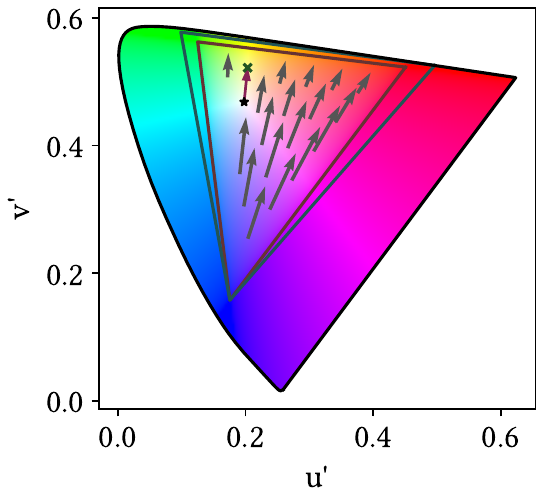}
    }
    \hfill
    \subfloat[sRGB color frequency under natural scenes.]
    {
    \label{fig:color_histogram}
    \includegraphics[height=1.75in]{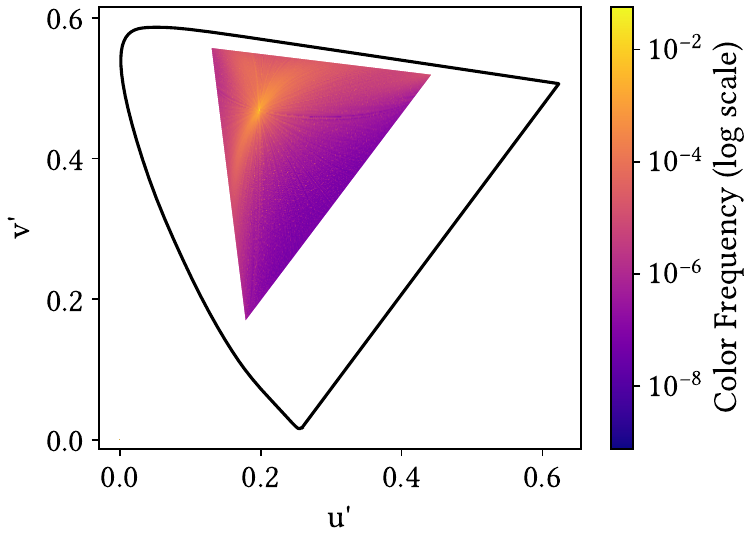}
    }
    \hfill
    \subfloat[Power-saving illuminants.]
    {
    \label{power:relative_power_by_illuminant}
    \includegraphics[height=1.75in]{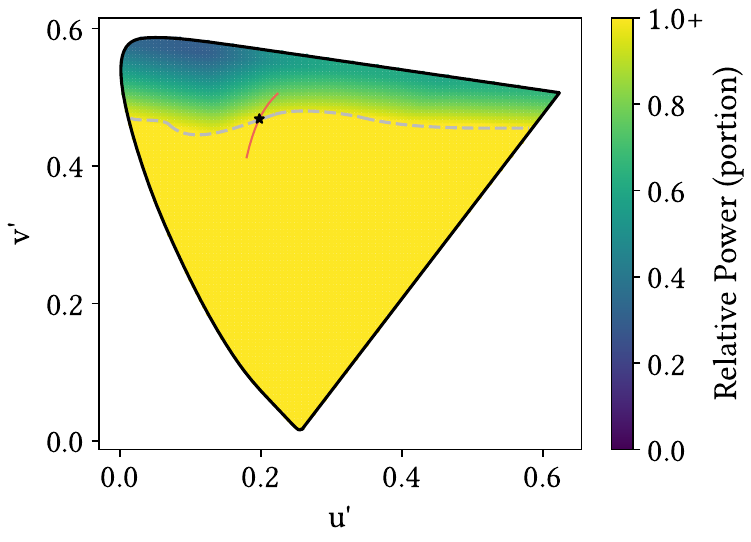}
    }
    \caption{(a) How sRGB colors are shifted under chromatic adaptation; when the illuminant shifts from \d65 to a yellow-ish illuminant, all colors shift toward yellow, too.
    (b) The distribution of colors in the natural scene dataset Places 365~\citep{places365} in CIE \upvp space.
    (c) Relative power consumption given the color distribution in (b) under different illuminants.  
    Illuminants are equiluminant and represented in the \upvp space.
    Power consumptions of each illuminant are normalized to that of \d65.
    The color map of the relative power consumptions is clipped at 1.
    All illuminants above the savings boundary yield an average reduction in power consumption.
    The daylight curve contains the CIE Standard Illuminant D series, representing the colors of daylights.
    }
\end{figure*}

Specifically, we obtain a color value distribution from a large natural scene dataset, Places365~\cite{places365}. 
\fixedme{
We use Places365 as opposed to more well-known datasets like ImageNet, as the images in Places365 are not of specific objects but, rather, of entire environments and, thus, are more true to VR viewing statistics.
\Fig{fig:color_histogram} shows the color density distribution from Places365 as a heatmap.
}

We then use each color's density to weight the power consumption of each color.
The form of average power saving is given in \Equ{eq:power_illu_a}, where $W_\texttt{c}$ is the weight given to color $\texttt{c}$, and $N = 256^3$ indicates that there are 16.7 million 8-bit sRGB colors; the weights are normalized such that $\sum_\texttt{c}^N W_\texttt{c} = 1$.
Our power estimation methodology can easily applied to a specific scene, if known, to obtain scene-specific power consumptions.

In summary, $\mP(\texttt{T})$ in \Equ{eq:power_illu} represents the power consumption of displaying an image, which is to be viewed under illuminant $\tT$ and has the same color appearance as that under the initial illuminant \d65, assuming an average color distribution of natural scenes.

\subsection{Results and Discussions}
\label{sec:illu:res}

We perform a sweep of illuminants $\tT$ within the human visual gamut and, for each $\tT$, estimate its power consumption using \Equ{eq:power_illu}.
The results are plotted in the CIE 1976 UCS chromaticity diagram, shown in \Fig{power:relative_power_by_illuminant}.
The power numbers are normalized to the power of the \d65 illuminant and are clipped at 1.
The colors in the figure represent the (relative) power consumption, with all the illuminant colors that lead to the same or higher power than that of \d65 plotted in yellow.
The gray dashed curve denotes the boundary that separates the power-saving illuminants and the power-increasing illuminants.
For reference, we also overlay on the plot 1) the daylight locus, which contains the colors of daylights~\citep{judd1964spectral} and 2) the gamut of the sRGB color space and the Display P3 color space.

We can see that greener and redder illuminants (corresponding to the top left and right of the \fixedme{\upvp} diagram) yield power savings.
This arises from a combination of two reasons.
First, chromatic adaptation theory dictates that shifting the illuminant towards red or green would generally shift other colors toward red and green too; see \Fig{fig:color_shifts_adapt} for a visual example.
Second, the blue subpixels in the particular OLED we use~\citep{chen2024pea, duinkharjav2022color} consume twice as much power as the red and green subpixels.
Thus, shifting colors toward red/green inherently reduces the strength of the blue channel, leading to power reduction.

\section{Pilot Study: Measuring and Modeling the Time Course of Chromatic Adaptation}
\label{sec:exp}

\begin{figure*}[h] 
    \centering
    \includegraphics[width=\linewidth]{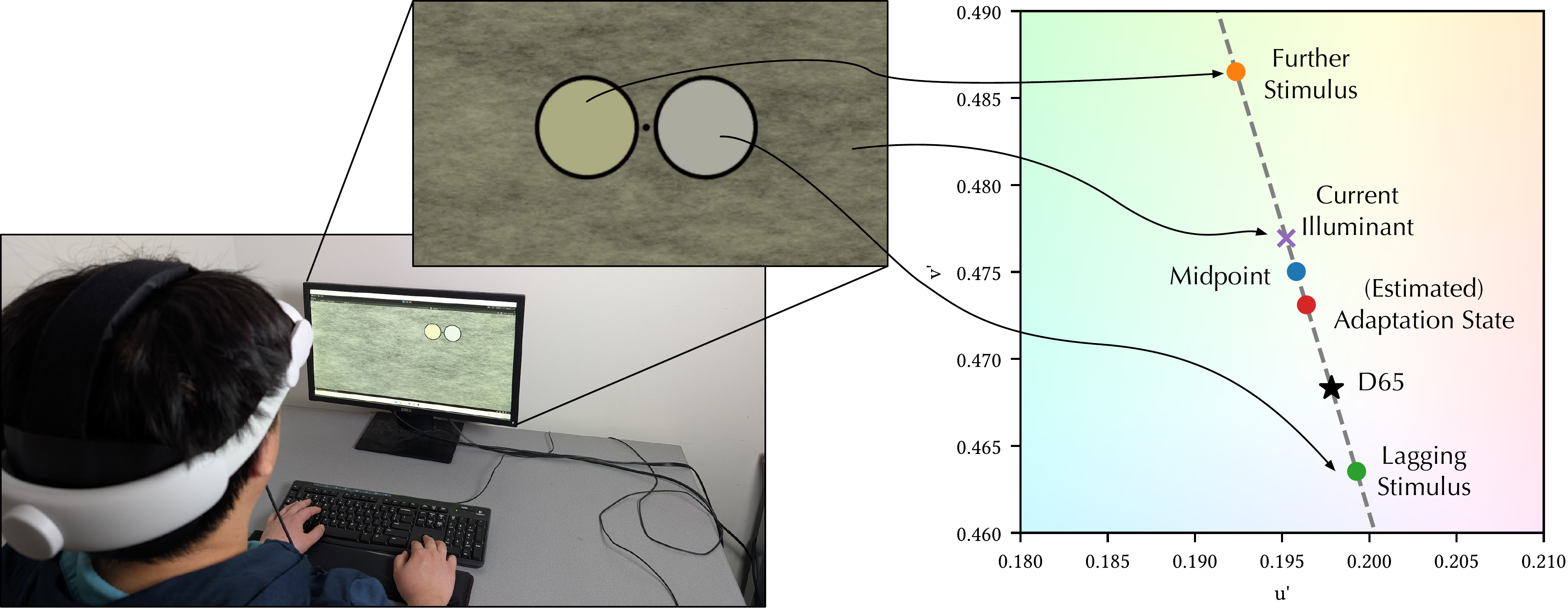}
    \caption{A visual example of the 2AFC task.
    Two colored stimuli on top of a background illuminant are flashed on the screen for 0.75 seconds, and the participant is asked to choose which patch is more achromatic.
    The background illuminant is a pink ($1/f$) noise pattern mimicking the natural scene statistics~\citep{ruderman1993statistics, field1987relations} and, in this case, has a green-ish average color as it shifts away from \d65.
    The ordering of the further stimulus (farther away from \d65) and the lagging stimulus (closer to \d65) is randomized across trials.}
    \label{fig:2afc_image}
    \label{fig:2afc_example}
\end{figure*}

\subsection{Goal and Basic Method}
\label{sec:exp:method}

Recall our ultimate goal: gradually change the initial illuminant of the scene $A(0)$, e.g., \d65, to a target illuminant to save display power consumption.
To minimize noticeable artifacts, the illuminant at any given time $t$, denoted $A(t)$, cannot be too great a distance from the user's adaptation state $a(t)$.
For instance, an instant change from $A(0)$ to a very chromatic illuminant does not afford the time one needs for chromatic adaptation and, thus, inevitably introduces noticeable artifacts.

The key challenge is, thus, to have a model that is capable of estimating the user's adaptation state $a(t)$ as the scene illuminant $A(t)$ changes over time $t$.
Inspired by \citet{time_course_chromatic}, we use a model that extends the classic Weber's law in psychophysics~\citep{fechner1948elements} and is analytically expressed as:
\begin{equation}
    a'(t) = \frac{\text{d}a(t)}{\text{d}t} = k_1 ( k_2 A(t) - a(t))
    \label{eq:adaptation_model}
\end{equation}

\fixedme{\Equ{eq:adaptation_model} models the rate of change of the viewer's adaptation state.}
The basic intuition is that chromatic adaptation takes place because what we think is white (i.e., our adaptation state or internal white point) is different from what the scene illuminant appears to be (i.e., the actual white point).
We hypothesize that the rate of adaptation ($a'(t)$) is proportional to the difference between the user's current adaptation state and the actual illuminant: when the illuminant appears more chromatic, we adapt faster, and vice versa.
Therefore, the first-order derivative of the adaptation function $a(t)$ is linearly correlated with $A(t) - a(t)$ up to a positive constant $k_1$.

It is known that humans can never fully adapt to extremely chromatic illuminants (\Sect{sec:bck}).
This is called ``partial adaptation.''
We introduce another parameter $k_2$ in \Equ{eq:adaptation_model} as the coefficient of adaptation completeness to account for partial adaptation.
The way to interpret $k_2$ is to consider the following scenario, where $A(t)$ is a constant for any $t>0$.
In this case, when $t$ approaches $\infty$, $a'(t)$ would approach 0, indicating the fact that given enough time our adaptation state would eventually settle under a constant illuminant.
When $k_2$ is 1, $a(\infty)$ must equal $A(\infty)$, indicating complete adaptation.
When $k_2 < 1$, $a(\infty) < A(\infty)$, indicating partial adaptation.


\Equ{eq:adaptation_model} is a standard first-order ordinary differential equation that can be solved if we know $k_1$ and $k_2$.
We regress $k_1$ and $k_2$ from pairs of $A(t)$ and $a(t)$ observations.
While we have control over $A(t)$, the challenge is to estimate $a(t)$, a participant's adaptation state at any given time $t$.
In an ideal world, we would ``freeze the time'' and probe the participant (e.g., adjust a test patch until the test patch appears achromatic).
In reality, however, the adaptation state of the participant would almost definitely shift while measuring.

Instead, our method is based on Maximum Likelihood Estimation (MLE).
We devise a psychophysical task that generates adaptation-state--dependent observations from participants, and then estimate the most probable adaptation state (at any given time) that maximizes the likelihood of the participants' responses being observed.

\subsection{The Maximum Likelihood Estimation Framework}
\label{sec:exp:mle}

To probe a participant's adaptation state, the core component of our psychophysics is a Two-Alternative Forced Choice (2AFC) task, where we present to participants two colored stimuli simultaneously against a background illuminant and ask which stimulus is less saturated, i.e., more achromatic.
An example of the task is shown in \Fig{fig:2afc_image}.
The colors of the two stimuli are a set distance apart in \upvp space, and are placed along the trajectory of the illuminant.
One stimulus' color is further ahead of the other stimulus' color in the direction of the illuminant's travel.
We term this patch the ``further'' patch, and the patch with the lagging color the ``lagging'' patch.

\begin{figure*}[h]
    \centering
    \begin{subfigure}[b]{0.49\textwidth}
        \centering
        \includegraphics[width=\linewidth]{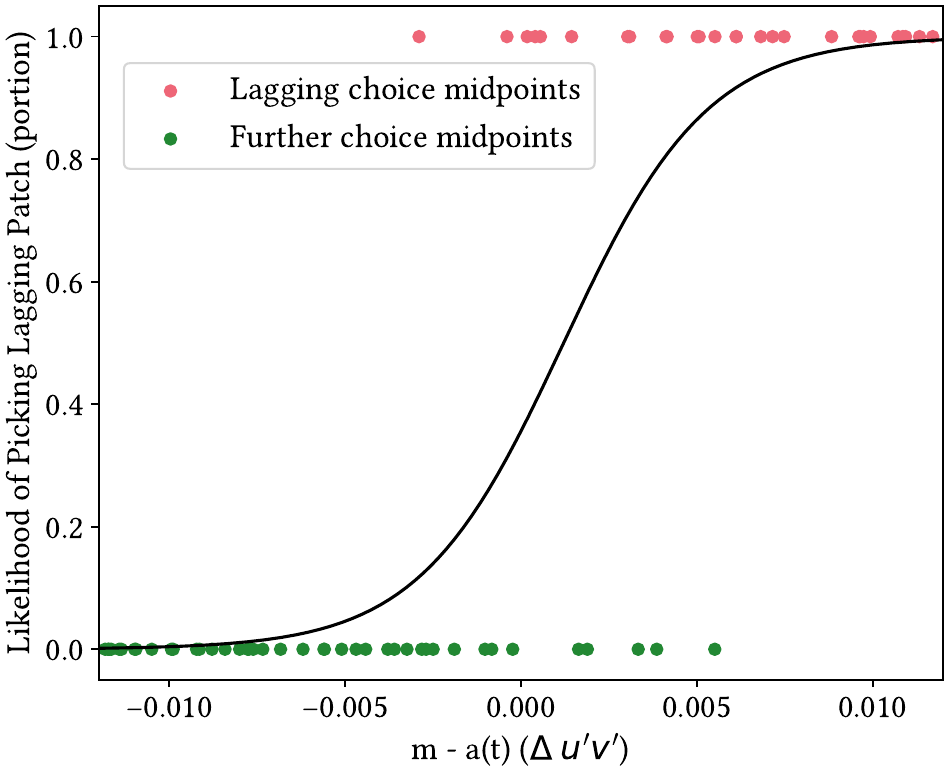}
        \caption{Raw data of an example calibration trial.}
        \label{fig:calibration_example}
    \end{subfigure}%
    \hfill%
    \begin{subfigure}[b]{0.49\textwidth}
        \centering
        \includegraphics[width=3.364431853in]{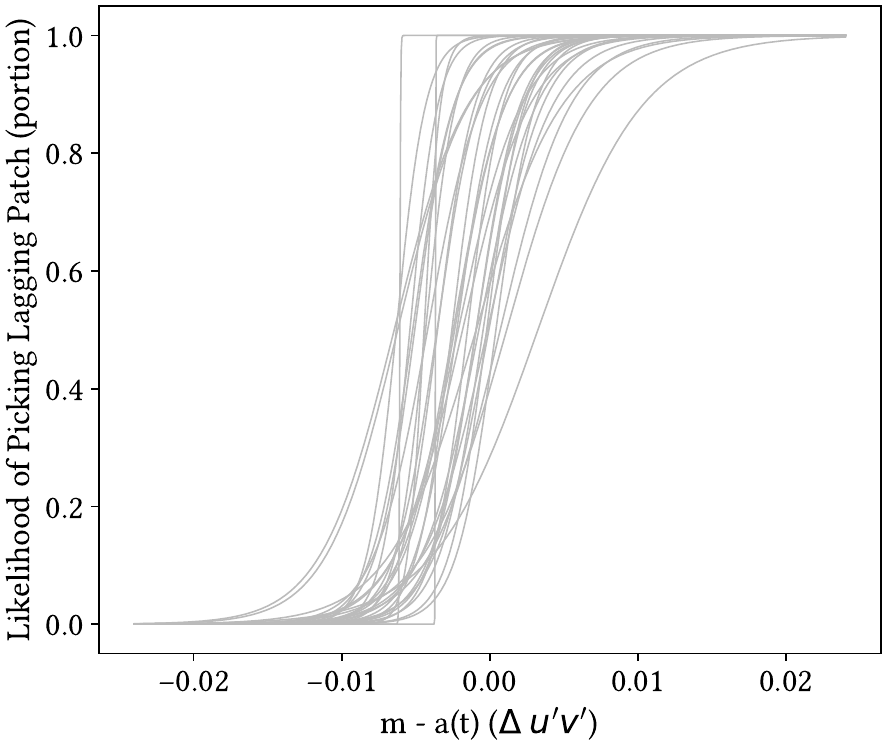}
        \caption{Aggregated psychometric functions across participants.}
        \label{fig:calibration_results}
    \end{subfigure}
    \caption{Results from the calibration stage.
    (a): Raw data from a calibration trial; pink/green markers represent the midpoint of a trial where a lagging/further stimulus is chosen.
    We then estimate a logistic psychometric function, representing the probability of a user picking the lagging patch, given the distance between the midpoint of the patch and the user's adaptation state.
    It is mathematically described in \Equ{eq:calibration_eq}.
    (b): Psychometric functions aggregated across participants and across sessions.}
    \label{fig:psychometric_function}
\end{figure*}

The background has a pink noise pattern to mimic the power spectrum statistics of natural scenes~\citep{ruderman1993statistics, field1987relations}.
The average color of the illuminant in \Fig{fig:2afc_image} gradually shifts from \d65 to a greener illuminant.
Therefore, the further patch is the green patch on the left, and the lagging patch is the gray patch on the right.

Informally, if $a(t)$, the participant's adaptation state at time $t$, is closer to the further patch, the participant is more likely to pick the further patch as being more achromatic; in contrast, if $a(t)$ is closer to the lagging patch, the participant is more likely to pick the lagging patch as appearing achromatic.
Let us use $m$ to denote the midpoint between the lagging patch $L$ and the further patch $F$, i.e., $m = \frac{F+G}{2}$ in the $u'v'$ space.
We define $G(m - a; \theta)$ as the \textbf{psychometric function} that quantifies the probability of picking the \textit{lagging} patch given the distance between $m$ and $a$;
$G(\cdot)$ is parameterized by $\theta$.

During our psychophysical study (whose protocol will be detailed in \Sect{sec:exp:proc}), we present a series of 2AFC tasks to the participants as the illumination shifts over time.
Each instance of the task at time $t$ has a different background illuminant $A(t)$, lagging patch $L(t)$, and further patch $F(t)$.
We use MLE to estimate the adaptation state $a(t)$ at any time $t$ by maximizing the likelihood function.
That is:
\begin{equation}
	\argmax_{k_1, k_2} \prod_{t\in \mathcal{L}} G(m(t)-a(t; k_1, k_2); \theta) \cdot \prod_{t\in \mathcal{F}} 1-G(m(t)-a(t; k_1, k_2); \theta),
    \label{eq:mle_basic}
\end{equation}

\noindent where we use $a(t; k_1, k_2)$ to make explicit that $a(t)$ is parameterized by the (to-be-estimated) $k_1$ and $k_2$ coefficients;
$\mathcal{L}$ is the set of all timesteps where a lagging stimulus is chosen, and $\mathcal{F}$ is the set of all timesteps where a further stimulus is chosen.
The likelihood function essentially represents the probability of observing the responses from a participant, where individual responses are governed by the probability distribution given by the psychometric function.

Two questions remain.
First, we need a way to parameterize and (experimentally) obtain the psychometric function $G(\cdot)$; this will be discussed in \Sect{sec:exp:cal}.
Second, we need a procedure to determine what exact $A(t)$, $L(t)$, and $F(t)$ to present to a participant at a given time $t$; this will be discussed in \Sect{sec:exp:proc}.
Then finally in \Sect{sec:exp:model} we will describe how to estimate $k_1$ and $k_2$ from the experimental data.

\subsection{Calibrating the Psychometric Function $G(\cdot)$}
\label{sec:exp:cal}

The psychometric function $G(\cdot)$ represents the probability of a user picking the lagging patch based on the distance between 
one's adaptation state and the middle point of the lagging and further patch.
We choose to parameterize $G(\cdot)$ using the logistic function commonly used in 2AFC psychophysics~\citep{wichmann2001psychometrici}, which has two free parameters $k$ and $x_0$ to be estimated:
\begin{align}
    G(m-a; k, x_0) = \frac{1}{1 + e^{-k(m-a-x_0)}}.
    \label{eq:calibration_eq}
\end{align}

To estimate $k$ and $x_0$, each participant goes through a calibration stage.
The participant enters the calibration stage fully adapted to \d65 by being exposed to a \d65 illuminant for 5 minutes, which is shown to be sufficient to fully adapt humans to \d65~\citep{fairchild2020kries, color_appearance_models}.
They then wear a VR headset where the calibration tasks is performed.
The background illumination is held constant at \d65 throughout the calibration stage , which means the adaptation state $a(t)$ is always \d65 during calibration, too.

We query the participant with 70 instances of the 2AFC task while varying the lagging and further stimuli in each instance.
The distance between the lagging and the further stimuli is always held at 6 Just Noticeable Differences (JND) in the \upvp space (1 JND is about 0.004 in \upvp space~\citep{wyszecki2000color}), but the midpoint $m(t)$ is normally placed between \d65 $\pm$ 3 JND.
\fixedme{We choose to separate the patches by 6 JND.
Separating the patches by lower distances increases the difficulty of the task while separating them by too much puts patches at risk of touching the display gamut.}
In each instance, the two stimuli are shown for 750 ms, followed by a 3 second interval during which the participant can input their answer.

An example of the calibration raw data is shown in \Fig{fig:calibration_example}, where the $x$-axis shows $m(t)-a(t)$ and the $y$-axis is the proportion picking the lagging patch.
Note that for each $m(t)-a(t)$, we poll the participant only once, so the proportion in the raw responses is always 0 (green; further stimulus chosen) or 1 (purple; lagging stimulus chosen)\footnote{
We find that polling each participant multiple times for a given midpoint yields similar psychometric functions but would dramatically increase the (already-long) total time for each participant (10 hours; see \Sect{sec:exp:proc}).
}.
Using the results of these responses, we can fit the $k$ and $x_0$ coefficients of the psychometric function using, again, MLE.
The solid curve in \Fig{fig:calibration_example} shows the psychometric curve fit to the individual responses of this example.

\Fig{fig:calibration_results} shows individual psychometric functions collected across participants and across study sessions.
It is interesting to observe that in many cases, the chance level of picking the lagging stimulus (0.5 on the $y$-axis) does \textit{not} correspond to $m(t)-a(t)=0$ on the $x$-axis.
That is, users have an internal bias toward which color is less achromatic even when presented with two stimuli that are equally distant from a color they regard as white (adaptation state).
Our psychometric functions capture this bias.

\subsection{Psychophysical Protocol}
\label{sec:exp:proc}

With the understanding of the psychometric function, we will now describe the protocol of our psychophysical study, which generates experimental data to estimate how the adaptation state $a(t)$ changes under a time-varying illuminant $A(t)$ (\Equ{eq:adaptation_model}).

We recruit $N=9$ participants (ages 18-25; 2 female, 7 male).
All participants have normal or correct-to-normal vision, and are all unaware of our study.
Each participant spends roughly 10 hours in total for the study, and the time is separated into five sessions that are at least one day apart.
We cap the duration per session at two hours to minimize lapse of concentration: we notice that the quality of the responses starts degrading after two hours.
All the experiments are approved by our IRB.

\begin{figure*}[ht]
    \includegraphics[width=2\columnwidth]{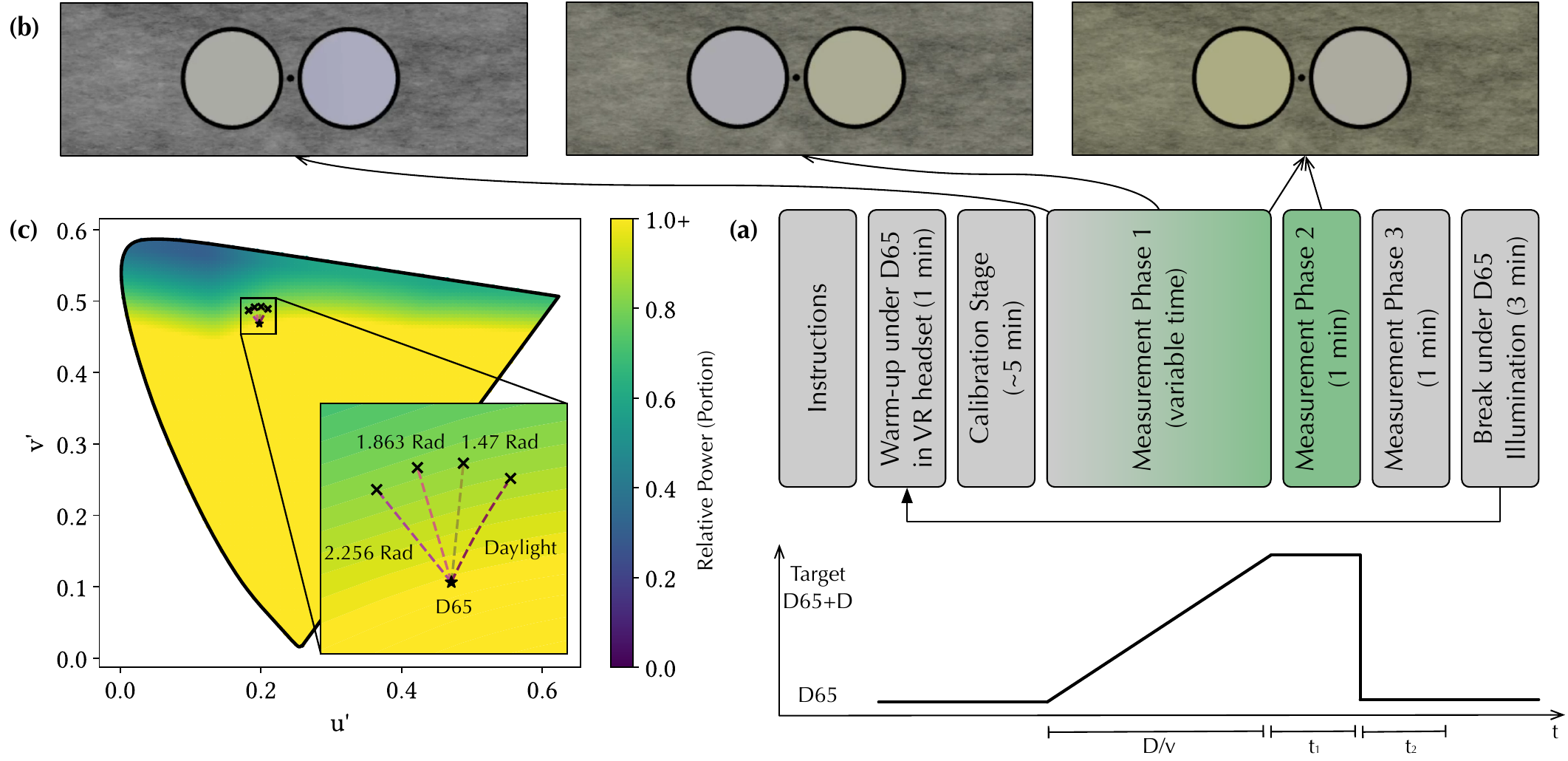}
    \caption{Structure of our pilot study.
    (a): The protocol of out psychophysical pilot study, where the illuminant shifts along a controlled trajectory and participants are asked to pick the more achromatic color patch of the two presented.
    (b): Examples of tests presented at the beginning, middle, and the end of the first measurement phase.
    (c): Illuminant trajectories applied in the first measurement phase; plotted on top of the relative power consumption by illuminant (normalized to D65 power clipped at 1).
    The trajectories are plotted as dotted colored lines.
    The terminal illuminant of each trajectory is indicated by a cross.
    }
    \label{pilot:structure}
\end{figure*}

\Fig{pilot:structure}\fixedme{a} illustrates the structure of our study.
We use a Meta Quest 3 VR headset for trials in our study.
The participants interact with our study through a keyboard.
We begin by administering a color blindness test and measuring the participant's IPD.
The participant then wears the headset, and goes through a training program that instructs the participant on how to perform the 2AFC psychophysical task.
Afterwards, we begin to run individual trials.

\paragraph{Trial Structure.}
Each trial is composed of a warm-up, calibration, and measurement stage.
\begin{itemize}
	\item In the warm-up stage, the participant is exposed to one minute of a background illuminant (without the test stimuli) with an average color of \d65.
	\item Following the warm-up stage is the calibration stage, which is used to derive the participant-specific psychometric function with a protocol described in \Sect{sec:exp:cal}.
	\item Next in the measurement stage, we gradually change the background illuminant (away from \d65) and the two stimuli, and probe the participants with a series of 2AFC tasks in order to estimate the adaptation state.
    The details of the measurement stage is discussed shortly.
\end{itemize}

At the end of the trial, the user takes off their headset for three minutes.
The test room is illuminated by a diffuse lamp with a correlated color temperature of 6500 K (closely matching a \d65 illuminant) and a color rendering index of 80.
This ensures that the participant's adaptation state slowly recovers to D65 over the course of the break.
Following the break, the participant wears the headset again and returns to the warm-up stage for another trial.

\begin{figure*}[h]
    \centering
    \begin{subfigure}[b]{\textwidth}
        \centering
        \includegraphics[width=\textwidth]{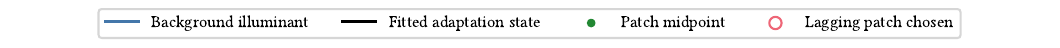}
    \end{subfigure} \\
    \centering
    \begin{subfigure}[b]{0.49\textwidth}
        \centering
        \includegraphics[width=\textwidth]{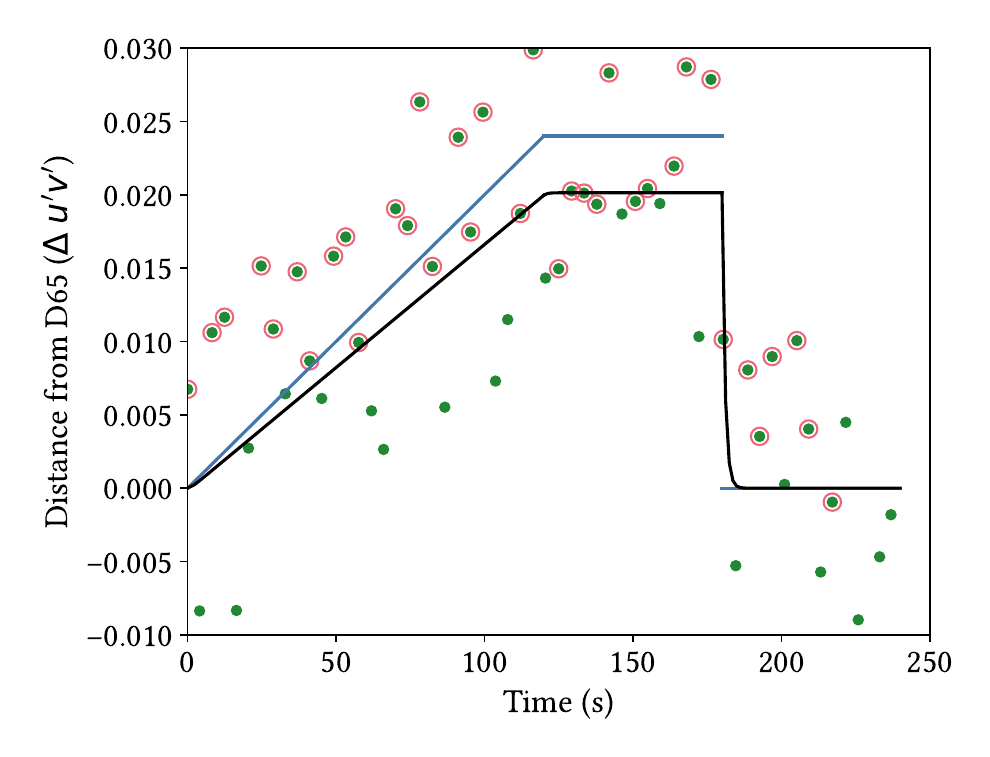}
        \caption{The illuminant traverses the daylight curve in the first phase.}
        \label{pilot:calibration_example_a}
    \end{subfigure}
    \hfill
    \begin{subfigure}[b]{0.49\textwidth}
        \centering
        \includegraphics[width=\textwidth]{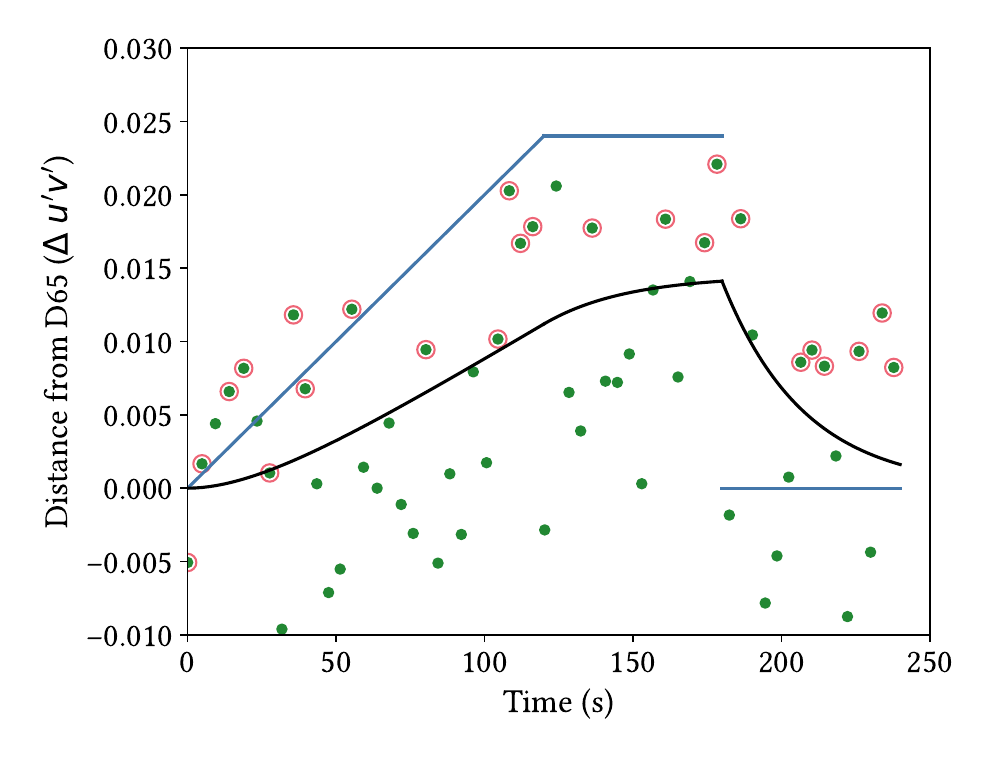}
        \caption{The illuminant traverses linearly toward 1.47 radians in the first phase.}
        \label{pilot:measurement_example_b}
    \end{subfigure}
    \caption{
    Two example trials of the triphasic measurement stage.
    The blue line represents the illuminant trajectory, and the black line represents the estimated adaptation state (through MLE).
    The green markers represent the midpoints of the two stimuli presented in the 2AFC tasks.
    If the lagging patch is chosen in a 2AFC response, the corresponding marker is circled.
    The adaptation state is much closer to the illuminant when the illuminant traverses the daylight locus (a) than the 1.47 radian trajectory (b), because humans are better at adapting to daylights~\citep{daylight_better, radonjic_daylight}.
    }
    \label{fig:example_composite}
\end{figure*}

\paragraph{Measurement Stage.}
The measurement stage is where we get the bulk of the data to estimate the adaptation state.
There are three phases in the measurement stage.

In the first phase, the background illuminant slowly traverses away from D65 at a set velocity $v$ along a direction $\phi$ in \upvp space.
The first phase ends when the illuminant reaches the ``terminal illuminant'', which is a distance D away from \d65.
In our study, we choose D to be 6 JNDs.
Each trial has a different traversal speed $v$ and a traversal direction $\phi$.
In the second phase, the illuminant is held at the terminal illuminant for a period of $t_1$.
In the third phase, the illuminant jumps back down to \d65 and stays there for a period of $t_2$.
\fixedme{To reduce the time consumed per trial, $D$ is limited to 6 JND and $t_1$ and $t_2$ are both set to one minute.}

As the background illuminant changes, we continuously present 2AFC tasks like the one shown in \Fig{fig:2afc_image}.
The two stimuli are always 6 JND apart, but the midpoint of the two stimuli is normally distributed around the running prediction of the adaptation state\footnote{The running prediction uses the same MLE-based method for offline prediction, which we discuss in \Sect{sec:exp:model}.}, a typical strategy in adaptive psychophysics~\citep{leek2001adaptive}.
Three example tests at the beginning, middle, and the end of the first phase are shown in \Fig{pilot:structure}\fixedme{b}.
As the trial progresses, the illuminant shifts toward green, and so do the two stimuli.

Throughout the measurement stage, we hold the luminance of the background illuminant constant, and the two stimuli have the same luminance as the background illuminant.

\paragraph{Parameter Selection and Scale of Study.}
Each trial, a different velocity $v$ and trajectory $\phi$ pair is selected.
We test four different trajectories and three different velocities for each trajectory.
For each trajectory and velocity pair (of which there are 12), we run 3 trials, leading to 36 trials per participant. 
We selected three evenly spaced velocities (defined as the distance in the \upvp space per second): $0.0001$, $0.0002$, and $0.0003$.
The velocities that we tested are bottlenecked by total test duration (10 total hours per participant).
\fixedme{We use these three velocities, as we find that the change in illuminant for velocities over $0.0003$ are trivially detectable. We lower bound the tested velocities at $0.0001$ due to time constraints.
At $0.0001$, each trial already takes time in excess of 16 minutes.}

\Fig{pilot:structure}\fixedme{c} shows the four trajectories we select overlaid over the power-saving landscape in the \upvp space.
Three of the four are linear trajectories while the other is the daylight locus (i.e., CIE Standard Illuminant D series)~\citep{judd1964spectral}; all start from \d65.

The three linear trajectories each have an angle of 1.47, 1.863, and 2.256 radians respectively, where an angle of 0 radians is defined to correspond to the horizontal direction pointing to the right in the \upvp space.
The illuminant trajectory of 1.863 radians was selected as that trajectory yielded the greatest power savings at 4 JND away from \d65, as identified by our power analysis in \Sect{sec:illu}.
The other two linear trajectories, 1.47 and 2.256, are 22.5 degrees clockwise and counterclockwise from 1.863. Both of these linear trajectories are well within the power-saving regime.

We choose the daylight locus as an additional trajectory because the human visual system is better at (with an evolutionary advantage) adapting to daylight illuminants~\cite{daylight_better, radonjic_daylight}.
The daylight locus is not a straight line; it has a standardized trajectory calculated from the emission spectra of daylight~\citep{condit1964spectral, henderson1963spectral}, and we use a quadratic parameterization of the locus by \citet{judd1964spectral}.

\subsection{Estimating the Adaptation State}
\label{sec:exp:model}

Given the triphasic measurement protocol (\Fig{pilot:structure}), $A(t)$ is analytically expressed as a piece-wise linear function:
\footnote{While technically the daylight locus is not linear, the portion of the daylight curve we test (D65 to around D50) is approximately linear, as seen in \Fig{pilot:structure}(c)},

\begin{align}
\begin{aligned}
\label{eq:bigA}
	A(t) &= 
	\begin{cases}
		\bd65 + v\cdot t \cdot \mathbf{u}
		, & 0 \leq t < \frac{D}{v} \\
		\bd65 + D\cdot\mathbf{u}, & \frac{D}{v} \leq t < \frac{D}{v} + t_1 \\
		\bd65, & \frac{D}{v} + t_1 \leq t \leq \frac{D}{v} + t_1 + t_2 \\
	\end{cases},\\
    \mathbf{u} &= \begin{bmatrix}
      \cos(\phi) \\
      \sin(\phi)
    \end{bmatrix},
\end{aligned}
\end{align}

\noindent where $\bd65 \in \mathbb{R}^2$ represents the chromaticity of \d65 in the \upvp space (we keep the luminance of the illuminant constant throughout a trial so the lightness dimension of the illuminant is omitted), and $D$ is the traversal distance in the first phase.

Given the analytical form of $A(t)$, we can analytically solve the differential equation in \Equ{eq:adaptation_model} to derive $a(t)$:

\begin{align}
\begin{aligned}
\label{eq:smallA}
	&a(t) = \\
&\begin{cases}
\bd65 + 
\left(\frac{k_2 v}{k_1} e^{-tk_1} + k_2vt-\frac{k_2 v}{k_1} \right) \cdot \mathbf{u} & 0 \leq t \leq \frac{D}{v}
\\
\bd65 + k_2 \left(D-\frac{v e^{-k_1 t} \left(e^{\frac{D k_1}{v}}-1\right)}{k_1}\right) \mathbf{u} & \frac{D}{v} \leq t < \frac{D}{v} + t_1
\\
\bd65 + \frac{k_2 e^{-k_1 t} \left(e^{\frac{D k_1}{v}} \left(D k_1   e^{k_1 t_2}-v\right)+v\right)}{k_1} \cdot \mathbf{u} & \frac{D}{v} + t_1 \leq t \leq \frac{D}{v} + t_1 + t_2
\end{cases}
\end{aligned}
\end{align}

We can see that $a(t)$ is parameterized by/dependent on $k_1$ and $k_2$.
Plugging \Equ{eq:smallA} back to the MLE-based optimization problem in \Equ{eq:mle_basic} (where the psychometric function $G(\cdot)$ is defined in \Equ{eq:calibration_eq}), we can solve for the best-fit $k_1$ and $k_2$, from which we obtain an estimate of $a(t)$ given $A(t)$.
This optimization problem is non-convex (due to the logistic psychometric function), and we use a fine-grained grid search (0.0004 increments along $k_1$, 0.0006 increments along $k_2$).
Solvers like gradient descent give worse results.
While perhaps obvious, it is worth emphasizing that the estimated $a(t)$ depends on the exact form of $A(t)$: a different equation for $A(t)$ would lead to a different equation for $a(t)$.



\subsection{Results and Discussion}

\begin{table}
    \caption{Best-fit $k_1$ and $k_2$ constants by illuminant trajectory. The trajectories all start from \d65, and are visualized in \Fig{pilot:structure}\fixedme{c}.}
    \centering
    \begin{tabular}{c|c|c}
        \toprule[0.10em]
         $A(t)$ trajectory (first phase) & $k_1$ & $k_2$ \\
         \midrule[0.05em]
         Daylight locus & 0.127 & 0.712\\ 
         linear @ 1.470 radians & 0.101 & 0.685 \\ 
         linear @ 1.863 radians & 0.107 & 0.638 \\
         linear @ 2.256 radians & 0.069 & 0.707 \\
         \bottomrule[0.10em]
    \end{tabular}
    \label{pilot_results:k1k2}
\end{table}

\Fig{fig:example_composite} shows two example trials, one for the daylight trajectory and the other for the 1.47 radian trajectory in the first phase.
In each plot, the $y$-axis represents the distance between the current illuminant and \d65 in the \upvp space.
The blue line represents the trajectory of the illuminant $A(t)$, and the black line represents the estimated adaptation state $a(t)$.
The green markers represent the midpoints of the two stimuli presented in the 2AFC tasks.
If the lagging patch is chosen in a 2AFC response, the corresponding marker is circled.

The trend of $a(t)$ is intuitive given the triphasic $A(t)$.
$a(t)$ initially rises as $A(t)$ rises in the first phase, indicating that chromatic adaptation is taking place (i.e., more chromatic colors are regarded by participants as ``white'').
$a(t)$ then gradually saturates as $A(t)$ is held constant at the terminal color in the second phase, and finally decays as $A(t)$ jumps back to \d65.

The $a(t)$ curve actually justifies our choice of a triphasic measurement protocol, which better estimates $k_1$ and $k_2$ than if only the first phase was used.
We can see that $a(t)$ in the first phase is close to linear\footnote{Technically the adaptation \textit{slightly} accelerates during this phase, as evidenced by the slightly increasing slope of $a(t)$; this is because $a(t)$ always lags behind $A(t)$ and the gap grows over time.}, which indicates just one free parameter could reasonably fit the curve.
As a result, the two parameters $k_1$ and $k_2$ could not be robustly estimated.
This is addressed by adding the second phase that decouples $k_1$ and $k_2$.
Examining the analytical solution of $a(t)$ in the second phase in \Equ{eq:smallA}, we can see that as $t$ increases, $a(t)$ approaches $k_2 D$, isolating the effect of $k_1$.

\fixedme{\Tbl{pilot_results:k1k2} summarizes the fits of $k_1$ and $k_2$ under the four trajectories, each estimated from the data across all velocities and across all nine participants.
\Apx{sec:uncertainty} shows the confidence intervals of these fits, which are narrow and are indicative of good fits.}
The two coefficients depend on the $A(t)$ trajectory, which is unsurprising considering their interpretations: $k_1$ represents the rate of adaptation and $k_2$ represents the adaptation completeness, both of which are affected by the illuminant.

For instance, the two coefficients are the highest for the daylight trajectory, consistent with previous literature that finds that the human visual system better adapts to daylight illuminants~\cite{daylight_better, radonjic_daylight}, so the rate of adaptation is faster and the adaptation is more complete,
which is captured by our psychophysical study and the estimation method.
This is also visually reflected in the two plots in \fixedme{\Fig{pilot:calibration_example_a} and \Fig{pilot:measurement_example_b}}, where the adaptation state under the daylight traversal is much closer to the illuminant and the slope is much higher than is under the 1.47 radiance trajectory.



\section{Chromatic Adaptation-Guided Power Optimizations}
\label{sec:power}

Using the adaptation model, this section describes a method that gradually shifts the illuminant in a virtual scene to reduce display power with minimal perceptual impacts.

\subsection{Formulation}
\label{sec:power:formulation}

To save power, the illuminant should gradually shift toward a power-saving color, as discussed in \Sect{sec:illu} and shown in \Fig{power:relative_power_by_illuminant}.
The illuminant should be shifted slowly to allow users to adapt, but a VR user spends only a finite amount of time in a session, so an overly slow shift reduces the potential power savings and is undesirable as well.

Our method, thus, imposes a time limit on the illuminant change --- the illuminant trajectory must arrive at the terminal illuminant within $t_{\max}$ seconds. 
The illuminant then remains constant for the rest of the usage session.
Our goal is to maximize the steady-state power saving (after the illumination stabilizes) while minimizing the perceptual impact.
We quantify the perceptual impact by the distance between the user's adaptation state, $a(t)$, and the illuminant, $A(t)$, in the \upvp space.
Therefore, the optimization problem is formally stated as:

\begin{align}
\begin{aligned}
    & \argmin_{A(t)} \mathcal{P}(A(t_{max}))\\
    & \text{s.t.:~~} \max_{0 \leq t\leq t_{\max}} [A(t) - a(t)] \leq \Delta T,
    \label{eq:optimization:opt_func}
\end{aligned}
\end{align}

\noindent where $\mathcal{P}(\texttt{T})$ is the average power consumption under an illuminant $\tT$ as discussed in \Sect{sec:illu:method}\footnote{We could also estimate the power frame by frame for a given video if known but for generality we choose the average power from natural scene statistics.}, and $\Delta T$ is the permissible difference between one's adaptation state and current illuminant.
The constraint is specified to ensure that the perceptual impact is within the $\Delta T$ allowance at all times during the illumination change; the adaptation state $a(t)$ will only get closer to $A(t)$ after the illuminant stabilizes, so no constraint need to be specified for $t > t_{max}$.

\subsection{Solution}
\label{sec:power:sol}

The optimization problem is too general to solve because $A(t)$ can technically take any arbitrary form (or even expressed only numerically): the illuminant can traverse an arbitrary curve in an arbitrarily time-varying speed.
To reasonably limit the scope of possible $A(t)$, we impose two simplifying constraints.
First, the illuminant must traverse one of the four curves we have experimentally measured in our pilot study (\Sect{sec:exp:proc} and \Tbl{pilot_results:k1k2}): three linear traversals plus the daylight curve.
Second, the traversal velocity is the constant that should be optimized.



Since there are only four traversal curves, we solve for the optimal velocity for each curve independently.
We make two observations.
First, $A(t) - a(t)$ monotonically increases, given that $A(t)$ follows a linear trajectory (see \Apx{sec:atat} for a proof).
Therefore, the constraint in \Equ{eq:optimization:opt_func} can be simplified to $A(t_{max}) - a(t_{max}) \leq \Delta T$.
Second, for a given traversal curve, traversing farther always leads to higher power savings (see \Fig{power:relative_power_by_illuminant} for an intuition).
Therefore, minimizing power consumption is equivalent to maximizing the traversal distance at $t_{max}$.

Given these two observations, the optimization formulation in \Equ{eq:optimization:opt_func} is simplified to (for a given trajectory):

\begin{align}
\begin{aligned}
    & \argmax_{v} A(t_{max}; v)\\
    & \text{s.t.:~~} A(t_{max}; v) - a(t_{max}; v) \leq \Delta T.
    \label{eq:optimization:opt_func:new}
\end{aligned}
\end{align}

Since both $A(t_{max}; v)$ and $A(t_{max}; v)-a(t_{max}; v)$ are monotonically increasing with $v$ (see \Apx{sec:atat}), the optimal $v$ that maximizes $A(t_{max})$ is when $A(t_{max}) - a(t_{max}) =\Delta T$ (the proof is a simple proof by contradiction, which we omit):
\begin{equation}
    v = - \frac{e^{k_1t_{max}} k_1 \Delta T}{k_2 - e^{k_1t_{max}} \left( k_2 + k_1 t_{max} -k_1k_2t_{max} \right)}
    \label{eq:optimization:analytical_velocity}
\end{equation}

\begin{figure}
    \includegraphics[width=\linewidth]{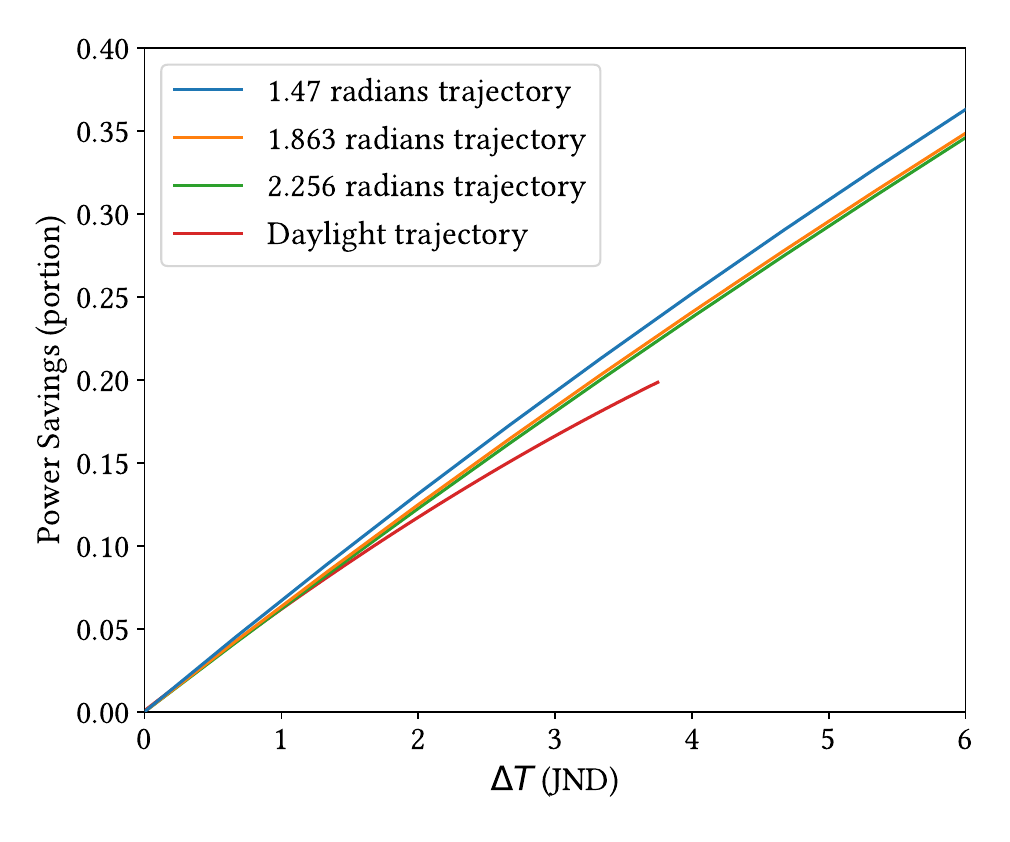}
    \caption{
    The tradeoff between power savings over \d65 ($y$-axis) and permissible perceptual loss expressed in units of JND ($x$-axis) under the four illuminant trajectories we study.
    Overall, the 1.47 radians trajectory Pareto-dominates the others.
    \fixedme{The daylight trajectory's power saving curve here is cut off short, as the CIE D-series illuminant is defined only over a limited distance.}
    Each trajectory at each $\Delta T$ has an underlying optimal $v$ (not shown) .
    }
    \label{fig:optimization:pareto}
\end{figure}

\begin{figure*}
    \centering
    \includegraphics[width=2\columnwidth]{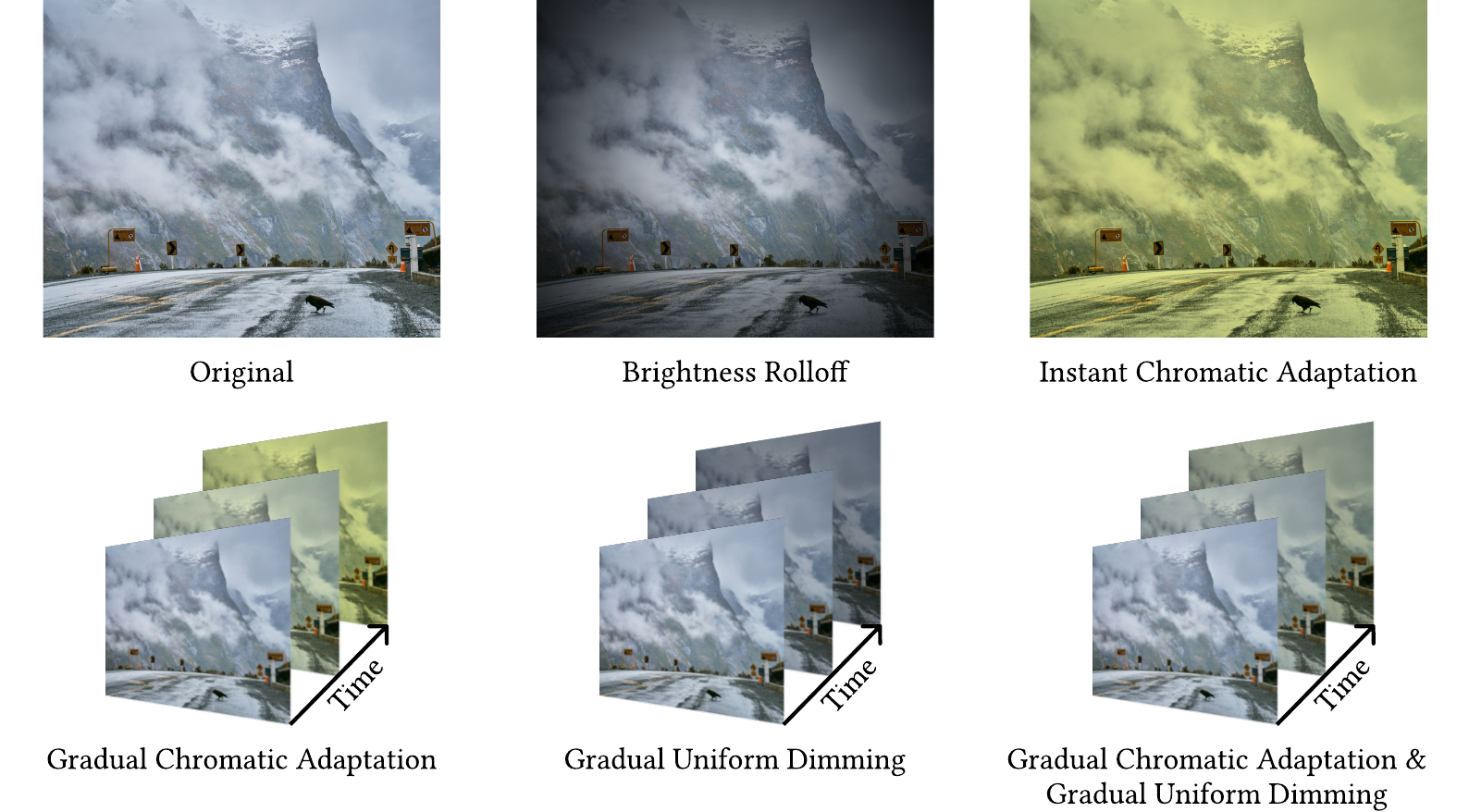}
    \caption{The various power-saving algorithms we benchmarked.
    Credit for the original image goes to Pedro Szekely~\cite{kea_borb}.}
    \label{fig:pilot:comparison}
\end{figure*}

\subsection{Results and Discussions}
\label{sec:power:res}

\Fig{fig:optimization:pareto} shows, for each traversal trajectory, how the power savings ($y$-axis) \fixedme{vary} with the permissible perceptual impact $\Delta T$ ($x$-axis), which is quantified in units of JND.
We set the time limit parameter $t_{\max}$ to be 2 minutes, which is much shorter than the average length of a VR session estimated to be around 38 minutes~\citep{alsop_vrar_2022}.
The power saving is calculated over the power consumption under the typical \d65 illuminant.
Unsurprisingly, as the allowance for perceptual impact $\Delta T$ increases, the power saving increases, too.

The linear traversal at $1.47$ radians is better than the other trajectories and, in particular, the daylight curve.
This is interesting because illuminants mostly studied in the literature and used in the real world are limited to those on the daylight curve --- for a good reason: evolutionarily humans are much better at adapting to daylights~\citep{daylight_better, radonjic_daylight}.
Our results show that adapting the illuminant along the 1.47 radians trajectory leads to lower perceptual loss under the same power consumption.

It might initially seem rather restrictive to study only four trajectories.
We note, however, that the fundamental psychophysical paradigm and the power optimization method are generalizable to other trajectories if one wishes.
Robustly interpolating or extrapolating to other trajectories would likely require more than four data points, which is challenging, given that each trajectory already takes over 10 hours per participant to collect data for.
Even with these restrictions, however, we are still able to identify power-efficient adapting illuminants that are previously unexplored.

\subsection{Implementing the Power-Saving Algorithm}
\label{sec:power:impl}

We implement a real-time illuminant shift algorithm on a Meta Quest Pro headset to demonstrate our power-saving method in practice.
The algorithm is implemented as a pair of programs --- a fragment shader written in ShaderLab and a director written in C\#.

The director controls the illuminant $A(t)$ at any given time.
$A(t)$ is calculated using the optimal 1.47 radians trajectory with a 0.000467 velocity in \upvp space calculated in \Sect{sec:power}.
The director performs interpolation between \d65 and the terminal illuminant across the 2 minute time course of the illuminant shift, and feeds the current illuminant to the fragment shader.
The fragment shader applies the CAT to all the image pixels given a particular $A(t)$ in each frame.
We use the commonly used linear Bradford CAT function \citep{lindbloom_cat_compare} as discussed in \Apx{sec:cat}.
We also show the pseudocode of the fragment shader in \Apx{sec:bradford}.


\paragraph{Computational Cost.}
The shader is lightweight to run.
Linear Bradford CAT amounts to a simple linear transformation per pixel in the linear sRGB space.
For a Meta Quest Pro headset, with two 1920 by 1800 pixel displays, our algorithm takes approximately 7.5 Giga Floating Point Operations (FLOPs) for rendering 90 FPS video.
For reference, the Quest Pro headset's GPU is capable of outputting 1.42 Tera FLOPs~\citep{adreno650}.
The director is even cheaper than the shader, only performing a handful of FLOPs per frame.
Our implementation uses less than 0.6\% of the overall compute capabilities of the mobile GPU and, thus, is unlikely to result in significant increase in rendering power consumption or latency.







\section{Validation Study}
\label{sec:validation}

\subsection{Baseline Methods}
\label{sec:validation:comparison}

Our method, henceforth termed Graduate Chromatic Adaptation (GCA), applies an illuminant shift in the first 2 minutes and then stabilizes the illuminant for the rest of the usage session, as described in \Sect{sec:power}.
We benchmark GCA against four state-of-the-art techniques, which are based on the experiments done in PEA-PODs~\cite{chen2024pea}.
PEA-PODs is a recent study that compares the perceptual impact of various power-saving rendering techniques for OLED displays.
\Fig{fig:pilot:comparison} presents a visual example comparing the various power-saving algorithms.

\paragraph{Instant Chromatic Adaptation (ICA).}
PEA-PODs assessed a chromatic-adaptation--based algorithm in their study, but their implementation did not account for the time course of chromatic adaptation; rather, the illuminant shift was applied instantly.
In particular, participants were free to switch between two images, one without the illuminant shift shader and the other with the shader turned on.
Between images, a black screen was presented for half a second.
The shader was immediately applied after the black screen, giving participant's visual system no time to adapt to the change in illuminant. 
We benchmark gradual chromatic adaptation against ICA to demonstrate the need for gradual illumination shift. 

\paragraph{Brightness Rolloff (BR).}
The second baseline we compare against is Brightness Rolloff, which gradually reduces the luminance of pixels the further they are from the fovea.
It is worth noting that that BR require gaze tracking, since it is inherently a gaze-contingent rendering method.
PEA-PODs observes that at 31\% power savings, BR results in the least perceptual impact compared to other methods --- when gaze tracking power is ignored, but becomes much less effective when the gaze tracking power is included.
In our implementation, we assume a gaze tracking power of 100 mW as assumed in prior work~\citep{chen2024pea}.

\paragraph{Gradual Uniform Dimming (GUD).}
Uniform dimming uniformly reduces the luminance of the visual field.
PEA-PODs' implementation applies the uniform dimming shader instantly, but uniform dimming relies on bleaching recovery adaptation, which, like chromatic adaptation, is gradual~\citep{barlow1972dark, hecht1937influence}, because both photoreceptor regeneration and re-sensitization (after being exposed to a stronger light) take time~\citep{fain2001adaptation, leibrock1998molecular, rushton1965ferrier, lamb2006phototransduction}.
We, thus, implement Gradual Uniform Dimming (GUD), which gradually dims the image over the time course of 2 minutes, which is the same duration as the illumination shifts in our method.

\paragraph{GUD + GCA.}
We also combine our gradual chromatic adaptation (GCA) algorithm with GUD, where GCA is applied over the course of the first minute and GUD is applied in the second minute.

\paragraph{Control.}
The control condition shows the original videos without any power-optimizing shader applied.
We observe that some participants report seeing artifacts in some original videos.
Including the control condition allows us to measure this inherent bias and analyze all other conditions against control.

\begin{figure*}[t]
\vspace{-2pt}
  \begin{minipage}[t]{0.7\columnwidth}
    \centering
    \includegraphics[height=1.7in]{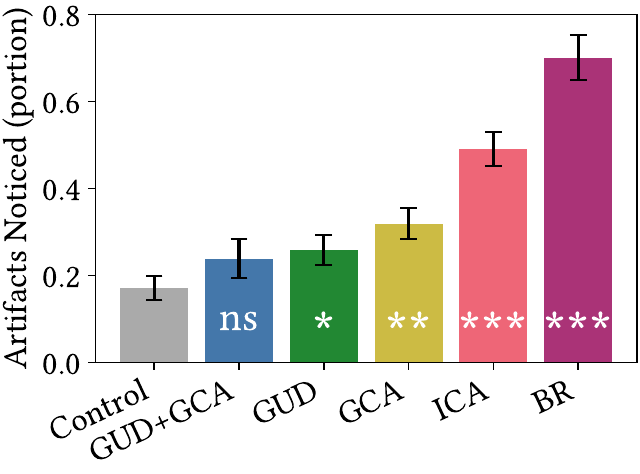}
    \caption{A comparison of participants' ability to notice artifacts in each condition.
    A star means statistically significantly different from control ($p < 0.05$);
    two stars indicate that $p < 0.01$, 
    three stars means that $p < 0.001$.
    GUD+GCA is the only method that is not statistically different (``ns'') from control.
    The error bars represent 1 standard error.}
    \label{fig:valid:compare_bar}
  \end{minipage}
  \hfill
  \begin{minipage}[t]{0.6\columnwidth}
    \centering
    \includegraphics[height=1.7in]{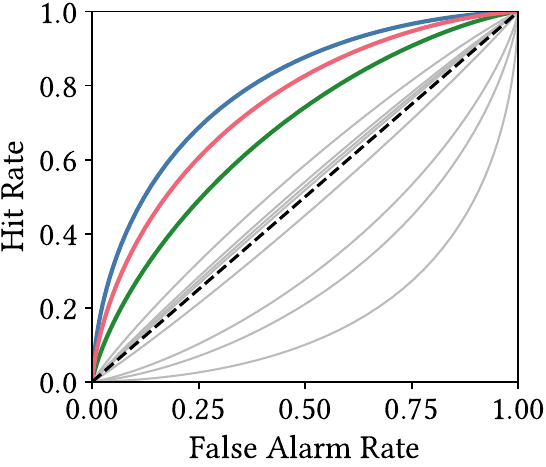}
    \caption{Receiver Operating Characteristic (ROC) curves for the GUD+GCA condition plotted by participants.
    Three participants (colored) are significantly better at detecting this condition than the others (hit rate much higher than false alarm rate).
    See texts for a discussion on potential improvements.}
    \label{fig:valid:roc}
  \end{minipage}
  \hfill
  \begin{minipage}[t]{0.7\columnwidth}
    \centering
    \includegraphics[height=1.7in]{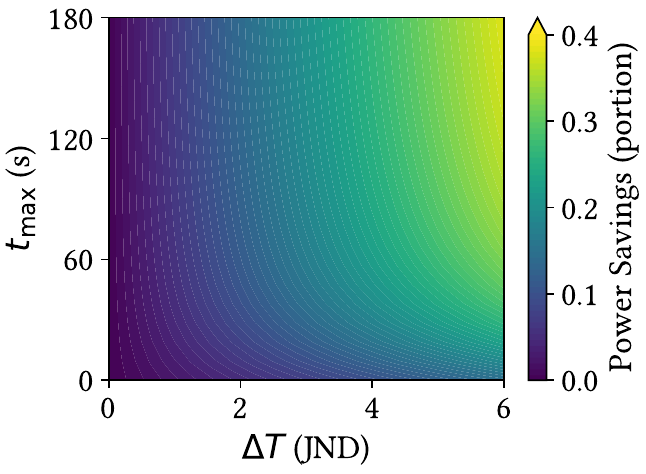}
    \caption{Heatmap of power savings as a function of both the permissible perceptual loss ($\Delta T$) and the time allocated for adaptation ($t_{max}$).
    Power saving is expressed with respect to the power consumption under \d65.
    We see diminishing returns in increasing $t_{max}$ and, to some extent, $\Delta T$.}
    \label{fig:validation:dt_vs_tmax}
  \end{minipage}
\end{figure*}

\subsection{Study Design}
\label{sec:validation:design}


\paragraph{Stimuli and Participants.}
The stimuli are five panoramic videos with neutral (\d65) in lighting and varied in content, each of which is clipped to 130 seconds in length.
Three of the videos were taken outdoors at approximately noon, one of the videos was taken from a video game, and one of the videos was taken indoors.

We recruit 20 participants (aged 18-27; 11 female) for the validation study;
none were aware of our research, nor did they participant in the pilot study in \Sect{sec:exp}.
All participants have normal or corrected-to-normal vision.
Participants used the Meta Quest Pro headset during the validation study\footnote{We used the Quest Pro here as opposed to the Quest 3 used in the pilot study in \Sect{sec:exp}, as we needed a headset with eye-tracking capabilities to benchmark the Brightness Rolloff algorithm.}, where the panoramic videos are rendered.
The participants perform tasks with a keyboard.

\paragraph{Structure.}
As discussed in \Sect{sec:power}, the power saving varies with the permissible perceptual loss.
We test a $\Delta T = 5$ JND threshold, which gives about 31\% power saving using our GCA algorithm.\footnote{\fixedme{The mean power savings for the five videos benchmarked in the validation study was even greater than the expected mean power savings, at 33\%.}}
We compare different algorithms by evaluating their perceptual impacts under the same power consumption budget.
The gaze-tracking power is deducted from the power budget for the BR method, the only gaze-contingent method among all methods.
We also test a 3 JND threshold, which gives about 20\% power savings for GCA. The results are detailed in \Apx{sec:3jnd}.

Three of the algorithms that we test (GCA, GUD, and GUD+GCA) are applied gradually.
Therefore, we cannot rely on users to compare pairs of stimuli using a Two-Interval Forced Choice (2IFC) paradigm, as done in \citet{chen2024pea} and \citet{ardavid}, as the gap between two stimuli would be too long (each video is 130 second long) for participants to reliably compare two stimuli.

Instead, we use a Two-Alternative Forced Choice (2AFC) paradigm similar to that in \citet{duinkharjav2022color}.
For each video we show a participant, we ask the participant if they perceive any visual artifacts in each video they watch (yes/no question).
Before the test begins, we show participants what artifacts they should expect to spot, and instruct them to only report those artifacts.

In total, each participant is presented with 60 randomized videos (5 scenes $\times$ 6 conditions $\times$ 2 repeats).
We cap the duration of each study to 2 hours per participant to ensure data quality, so not all participants complete all 60 trials.

\subsection{Results and Discussions}
\label{sec:validation:results}


\Fig{fig:valid:compare_bar} compares the chances of participants noticing the different conditions listed in \Sect{sec:validation:comparison}.
The error bars represent 1 standard error.
We perform statistical testing between control and each of the other methods using the Generalized Estimating Equation (GEE) technique~\citep{liang1986longitudinal, seabold2010statsmodels}, clustering data by participant and using the binomial distribution family.
We opt to use GEE, as 1) a participant's responses are correlated --- some participants have consistently higher or lower response biases (e.g. greater chance to report an artifact, even when none are present), 2) the number of trials between algorithms is unequal, and 3) we are interested in population-level differences in detectability. 

GUD+GCA results in the smallest chance of being noticed by participants over control.
In fact, the combination method is not statistically significantly different from control ($p = 0.213$), while all other methods are significantly greater than control ($p < 0.05$).
The fact that the combined method outperforms GUD and GCA alone suggests that these two techniques are complementary.
We discuss potential ways to further improve the combined method in \Sect{sec:disc}.

Gradual chromatic adaptation, unsurprisingly, outperforms the instant chromatic adaptation algorithm benchmarked in \citet{chen2024pea} by a significant margin ($p < 0.005$).
The difference in performance demonstrates the importance of gradually shifting the scene illuminant.
GCA is not statistically different from GUD ($p = 0.165$).
As the only gaze-contingent method, BR is the worst condition due to the high gaze-tracking power requirements (\Sect{sec:validation:design}).


\paragraph{Signal Detection Theory Perspective.}
We use signal detection theory (SDT) to offer another interpretation of the data.
The idea is to frame the validation study as detecting a signal (reporting a particular power-saving shader being applied) over noise (control condition).
We compute the discriminability index, $d'$ (d-prime), of each condition over control based on the hit rate (detecting a particular shader when it is applied) and false alarm rate (detecting a particular shader when it is \textit{not} applied)~\citep[Chpt. 6]{prins2016psychophysics}.
A similar analysis is performed by \citet{temporal_daylight} when analyzing their adaptation data.

\Tbl{tab:valid:dp} compares the $d'$ figures averaged across participants for each condition.
$d'=0$ means the hit rate equals the false alarm rate, indicating a chance-level performance.
A higher $d'$ means participants are better at detecting a shader being present.
Among all conditions, participants are least able to detect the GUD+GCA condition with an average $d'$ close to 0.
The ordering of $d'$ values matches that in \Fig{fig:valid:compare_bar}.
Bar charts comparing the mean $d'$ values of each algorithm and their standard deviations can be found in \Apx{sec:dp}.

\begin{table}[t]
    \centering
    \caption{The discriminability index $d'$ (closer to 0 is better) and z-scores for each of the different condition averaged across participants.}
    \renewcommand*{\arraystretch}{1}
    \renewcommand*{\tabcolsep}{5pt}
    \resizebox{.9\columnwidth}{!}
    {
        \begin{tabular}{c|c|c|c}
            \toprule[0.10em]
            \textbf{Method} & \textbf{Mean d'} & \textbf{Hit Rate} & \textbf{False Alarm Rate}\\
            \midrule[0.05em]
            GUD + GCA & 0.035 & 23.9\% & 20.6\% \\
            GUD & 0.209 & 25.9\% & 17.0\% \\
            GCA & 0.424 & 31.9\% & 17.0\% \\
            ICA & 0.909 & 49.1\% & 17.0\% \\
            BR  & 1.728 & 70.1\% & 12.5\% \\
            \bottomrule[0.10em]
        \end{tabular}
    }
    \label{tab:valid:dp}
\end{table}

\paragraph{Understanding Failure Cases.}
The GUD+GCA condition combines uniform dimming and gradual chromatic adaptation by simply cascading them.
It is encouraging that such a simple implementation is slightly, albeit not significantly, better than GUD or GCA alone.
To understand the ``failure cases'' under the combined condition, we use the SDT data to graph the Receiver Operating Characteristic (ROC) curve for every participant for the GUD+GCA condition in \Fig{fig:valid:roc}.
Additional ROC curves for different algorithms can be found in \Apx{sec:rocs}.

Three participants in particular have high ROC curves under GUD+GCA; they are significantly better at detecting the presence of the GUD+GCA shader compared to the other participants.
In contrast, two of three three participants have very low performance in detecting the GUD condition (not shown).

The performance of GUD+GCA can be improved in the future by addressing two limitations.
First, chromatic adaptation is luminance dependent~\citep{wei2019effects, vincent2016luminance}, but GUD+GCA uses a chromatic adaptation model tuned at one luminance level across all luminance levels during dimming.
An important future work is to extend our time course study to different luminance levels and apply a luminance-dependent illuminant shift.

Second, the adaptation model derived from the pilot study (\Sect{sec:exp}) may not apply to these three participants.
After all, the model is based on $k_1$ and $k_2$ coefficients regressed at the population level.
These three participants may be better suited through more personalized parameters, which we leave to future work.





\subsection{Power Saving Sensitivity}
\label{sec:validation:sen}

The results so far are based on the setting where adaptation time limit $t_{max}$ is 2 minutes (\Sect{sec:power:res}) and the permissible perceptual loss $\Delta T$ is 5 JND (\Sect{sec:validation:design}).
We analyze the sensitivity of power savings under varying $t_{max}$ and $\Delta T$.
\fixedme{The results of this analysis for the $1.47$ radians trajectory are shown in \Fig{fig:validation:dt_vs_tmax}.}
There are diminishing returns on power savings as $t_{max}$ increases.
This is because, at high time limits, the distance we are able to adapt the user to is limited more by the completeness of chromatic adaptation ($k_2$) than by the adaptation rate ($k_1$).
Sensitivity data under other trajectories can be found in \Apx{sec:tmax_dt_savings}.


\section{Limitations and Future Work}
\label{sec:disc}

\paragraph{Assumptions in Computational Modeling.}
We have assumed that the psychometric function $G(\cdot)$ in \Sect{sec:exp:cal} is a function of $m-a$, the distance between the adaptation state and midpoint of the two test stimuli, but not the absolute adaptation state $a$.
While this makes intuitive sense, future work could rigorously test this by performing the calibration test at under different illuminants.

In our computational model of the adaptation state (\Equ{eq:adaptation_model}),
we assume that $k_2$ is constant under a given trajectory, and is universal across all three phases of the pilot study (\Sect{sec:exp:proc}).
However, since the actual illuminant $A(t)$ changes over time, it would be more precise to parameterize $k_2$ as a function of $A(t)$.
The parameterization could be empirically derived by, for instance, individually measuring the adaptation completeness under varying illuminants.
There is also evidence that chromatic adaptation can even be affected by recent illuminant exposure history~\citep{fairchild2022reversibility, cai2018bidirectional}.
Therefore, the $k_1$ and $k_2$ coefficients might be modeled as a function of illuminant history over a time window.

\paragraph{Implementations.}
Our psychophysical pilot study, the computational model, and the validation study all assume that the original illuminant in a scene is \d65, which is a fair assumptions since \d65 is the white point of sRGB and Display P3, the two most commonly used color spaces in rendering and video content.
If the scene is rendered in a color space that uses a different white point, e.g., D60 in DCI-P3, one needs to specifically derive a D60-specific model --- using the our psychophysical paradigm and computational modeling methodology, which are generalizable to other illuminants.

Our adaptation model (\Sect{sec:exp}) enables a host of power optimizations.
Our current optimization (\Sect{sec:power}) is formulated as maximizing the power savings after the illuminant stabilizes.
Alternative formations exist.
For instance, one could maximize the total power saving throughout a usage session if the duration is known.
One could also minimize perceptual loss $A(t)-a(t)$ under a given power budget.

Our validation study assumes uninterrupted VR usage.
If a user takes off the VR headset and is exposed to the ambient illuminant, we have to restart the adaptation when they put the headset back on.
This would reduce the potential power saving opportunities.

\fixedme{
Our current implementation also assumes that the lighting in the original VR scene is relatively stable.
This assumption does not hold true for some VR content.
Our technique can, however, be readily extended to account for dynamic lighting conditions.
The exact formulation for this extension is detailed in \Apx{sec:dynamic}.
}

\fixedme{
\paragraph{Luminance Dependence.}
In principle, chromatic adaptation could be luminance dependent~\citep{wei2019effects, vincent2016luminance}, whereas the present study holds the luminance constant.
An important future work is to extend our study to different luminance levels and understand how the time course of chromatic adaptation is affected by luminance.
Such an extension would also inform a better implementation of the GCA + GUD algorithm, because GUD adjusts the scene luminance.
}

\fixedme{
\paragraph{Color Foveation.}
\citet{duinkharjav2022color} exploited the human visual system's reduced color discrimination in peripheral vision to shift rendered colors imperceptibly while reducing display power.
Their work builds a power model for OLED displays, which we use in this paper.
A key advantage of our technique compared to ~\citep{duinkharjav2022color} is that we do not rely on gaze tracking, which could be power hungry~\citep{chen2024pea, feng2024blisscam}.
However, color foveation and chromatic adaptation are not mutually exclusive.
To systematically exploit the two simultaneously, an interesting future work is to derive color discrimination thresholds under different illuminants and adaptation states.
}

\paragraph{Extension to Augmented Reality.}
While this study focuses on VR, chromatic adaptation is equally, if not more, important to AR, where color vision is constantly modulated by real-world lighting.
Accounting for chromatic adaptation is critical not only for saving power but also for accurate color reproduction.

The key for AR scenarios, of course, is illuminant estimation, which has long been studied in the context of auto white balancing and computational color constancy~\citep{gijsenij2011computational, barnard2002comparison, rowlands2020color}.
An interesting future work is to extend our psychophysics and modeling to AR and to integrate chromatic adaptation into AR perceptual models~\citep{ardavid}.
With lighting estimation become more mature in product AR toolkits~\citep{arkitlightingapi}, we expect techniques incorporating chromatic adaptation, whether for improving perceptual quality or for reducing power, to be increasingly practical and deployable in real-world AR devices.

\section{Conclusion}
\label{sec:conc}

This paper makes both a scientific contribution and an engineering contribution.
Scientifically, we propose a novel psychophysical paradigm along with the statistical inference method to model a user's adaptation state given the trajectory of dynamic illuminant change.
We then demonstrate an engineering application of this model.
We devise an algorithm to gradually shift the scene illuminant in virtual environments to reduce OLED display power consumption by 31\% with little perceptual impact.

\begin{acks}
The work is partially supported by NSF
Awards \#2225860, \#2126642, and \#2044963 as well as a Meta research grant.
\end{acks}

\bibliographystyle{ACM-Reference-Format}
\bibliography{refs.bib}

\newpage
\onecolumn
\setcounter{figure}{0}
\setcounter{equation}{0}
\appendix
\section{Background on Chromatic Adaptation Transform (CAT) Function}
\label{sec:cat}

Chromatic adaptation, phenomenologically, relies on scaling the sensitivities of the cone photoreceptors.
We discuss the key mathematical processes here, and refer interested readers to \citet[Chpt. 11]{macadam1982color}, \citet[Chpt. 9]{wandell1995foundations}, and \citet{color_appearance_models} for comprehensive discussions.
The cone responses of a color after scaling can be represented as:

\begin{align}
\label{eq:vk_scaling}
    \begin{bmatrix}
        1/L_\tS & 0 & 0 \\
        0 & 1/M_\tS & 0 \\
        0 & 0 & 1/S_\tS
    \end{bmatrix} \times
    \begin{bmatrix}
        L \\
        M \\
        S
    \end{bmatrix}_\tS,
\end{align}

\noindent where $L_\tS$, $M_\tS$, and $S_\tS$ represent the LMS cone responses of the white point/illuminant of the scene, and $[L, M, S]_\tS^T$ represents the unscaled LMS cone responses of an arbitrary color under illuminant $\tS$.
We can see that if $[L, M, S]_\tS^T$ \textit{is} the illuminant itself, the scaled cone responses would be constant (which is customarily normalized to $[1, 1, 1]^T$ in the literature) --- regardless of the color of the illuminant, which is discounted.

For color $\cS$ viewed under illuminant $\tS$ to have the same color appearance as $\cT$ viewed under $\tT$, the following must hold:

\begin{align}
\label{eq:cat}
    \begin{bmatrix}
        1/L_\tS & 0 & 0 \\
        0 & 1/M_\tS & 0 \\
        0 & 0 & 1/S_\tS
    \end{bmatrix} \times \cS = 
    \begin{bmatrix}
        1/L_\tT & 0 & 0 \\
        0 & 1/M_\tT & 0 \\
        0 & 0 & 1/S_\tT
    \end{bmatrix} \times \cT, 
\end{align}

\noindent where both $\cS\in\mathbb{R}^3$ and $\cT\in\mathbb{R}^3$ are expressed in the LMS cone space; $L_\tT$, $M_\tT$, and $S_\tT$ represent the LMS cone responses of the illuminant $\tT$.
That is, the scaled cone responses under both illuminants must match.

Usually colors are expressed in the (linear) sRGB space, in which case $\cT$ is transformed from $\cS$ by:

\begin{align}
\begin{aligned}
\label{eq:cat_srgb_supp}
    \cT = &f_{\tS\rightarrow\tT}(\cS) \\
    = &T_{sRGB2XYZ}^{-1} \times T_{XYZ2LMS}^{-1} \times 
    \begin{bmatrix}
        L_\tT/L_\tS & 0 & 0 \\
        0 & M_\tT/M_\tS & 0 \\
        0 & 0 & S_\tT/S_\tS
    \end{bmatrix}\times \\
    &T_{XYZ2LMS} \times T_{sRGB2XYZ} \times \cS,
\end{aligned}
\end{align}

\noindent where both $\cS$ and $\cT$ are expressed in the linear sRGB space, $T_{sRGB2XYZ}\in\mathbb{R}^{3\times3}$ is the transformation matrix from the linear sRGB space to the CIE 1931 XYZ space, and $T_{XYZ2LMS}\in\mathbb{R}^{3\times3}$ is the transformation matrix from the XYZ space to the LMS space.
$f_{\tS\rightarrow\tT}(\cdot)$ is the Chromatic Adaptation Transform (CAT) function (operating in the linear sRGB space here).
A commonly used $T_{XYZ2LMS}$ in chromatic adaptation literature is the Bradford's spectrally sharpened matrix~\citep{lindbloom_cat_compare}, with which the CAT function is called the linear Bradford chromatic adaptation (see \citet[Chpt. 15]{color_appearance_models} and \citet{luo1996llab}), which we use in this paper.

\section{Linear Bradford Pseudocode}
\label{sec:bradford}

\Alg{algo:linbrad} is pseudocode for the Linear Bradford algorithm~\citep{lindbloom_ca}.
The whitepoints are first converted into Linear Bradford space in lines 1 and 2.
On line 3, a diagonal matrix is computed that converts colors from the source white point to the target white point, in Linear Bradford space.
This matrix is converted back sRGB space, before being applied to the original image.

\begin{algorithm}
\caption{Linear Bradford}
\label{algo:linbrad}
\begin{algorithmic}[1]
\Statex \textbf{Input:} RGB image $\mathbf{I} \in \mathbb{R}^{H \times W \times 3}$, source white $\mathbf{W}_s \in \mathbb{R}^3$, destination white $\mathbf{W}_d \in \mathbb{R}^3$
\Statex \textbf{Constants:} Linear Bradford matrix $\mathbf{M}_B \in \mathbb{R}^{3\times 3}$
\Statex \textbf{Output:} Adapted image $\mathbf{I}' \in \mathbb{R}^{H\times W \times 3}$

\State $\mathbf{W}_s^{B} = \mathbf{W}_s \times \mathbf{M}_B$
\State $\mathbf{W}_d^{B} = \mathbf{W}_t \times \mathbf{M}_B$
\State $\mathbf{D}^B=\textrm{diag}(\mathbf{W}_t^{B} / \mathbf{W}_s^{B})$
\State $\mathbf{D}=\mathbf{M}_B\times\mathbf{D}\times\mathbf{M}_B^{-1}$
\State $\mathbf{I}'=\mathbf{I} \times \mathbf{D}$
\end{algorithmic}
\end{algorithm}
\section{Proof of $A(t) - a(t)$ monotonicity}
\label{sec:atat}

Assuming that the $A(t)$ trajectory follows a linear path, e.g., the first case in \Equ{eq:bigA} and \Equ{eq:smallA}, we first prove that $A(t)-a(t)$ monotonically increases with $t$.
$A(t) - a(t)$ is expressed as follows:

\begin{align}
    A(t) - a(t) &= vt - (\frac{k_2 v}{k_1} e^{-tk_1} + k_2vt - \frac{k_2 v}{k_1}), \\
    \frac{\partial(A(t) - a(t))}{\partial t} &= v + k_2 v e^{-tk_1} - k_2v = (1-k_2)v + k_2 v e^{-tk_1}.
\end{align}

$v$, $k_1$, and $k_2$ are always positive.
Therefore, the term $k_2 v e^{- t k_1}$ is always positive.
Additionally, $k_2$ represents the completeness of adaptation and is bounded between $0$ and $1$.
Therefore, $1 - k_2 \geq 0$.
The partial derive of $A(t) - a(t)$ is, thus, lower bounded by 0, and the function $A(t) - a(t)$ monotonically increases with $t$.

We then prove that $A(t_{max}; v)-a(t_{max}; v)$ monotonically increases with $v$.

\begin{align}
    \frac{\partial(A(t_{max}; v) - a(t_{max}; v))}{\partial v} &= t - \frac{k_2}{k_1} e^{-t_{max}k_1} - k_2t_{max} + \frac{k_2}{k_1} = (1-k_2)t_{max} + \frac{k_2}{k_1}(1-e^{-t_{max}k_1}) > 0.
\end{align}


\section{$t_{max}$ vs $\Delta T$ vs Power Savings by Trajectory}
\label{sec:tmax_dt_savings}

\begin{figure*}[h]
    \centering
    \subfloat[1.863 radians trajectory]
    {
        \label{fig:tmax_dt_savings:1863}
        \includegraphics[height=1.6in]{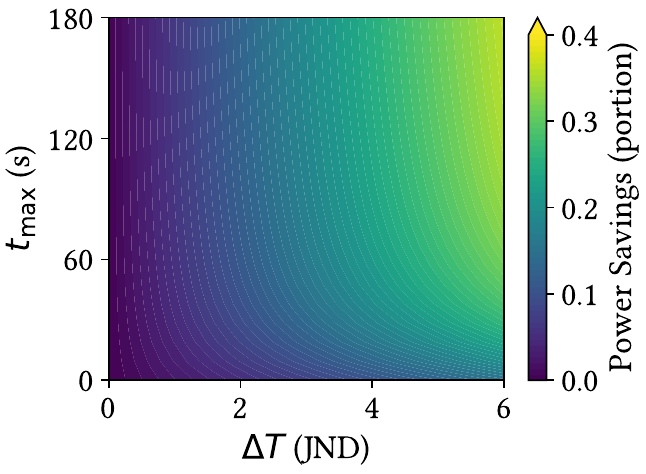}
    }%
    \subfloat[2.256 radians trajectory]
    {
        \label{fig:tmax_dt_savings:2256}
        \includegraphics[height=1.6in]{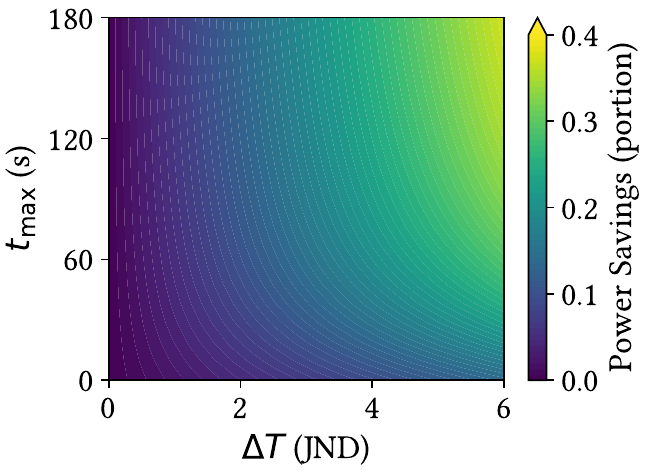}
    }%
    \subfloat[Daylight trajectory] {
        \label{fig:tmax_dt_savings:daylight}
        \includegraphics[height=1.6in]{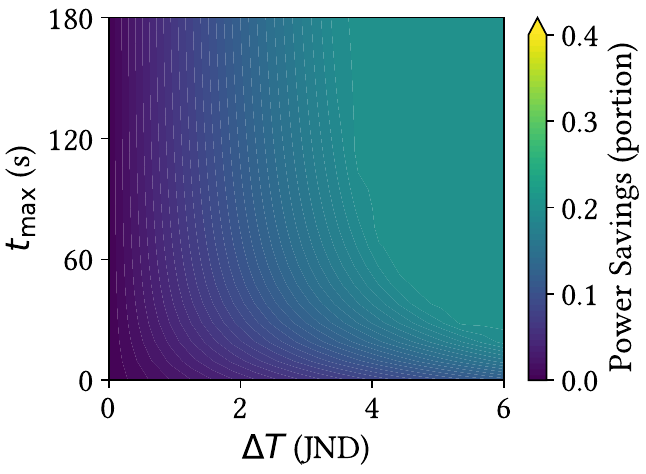}
    }%
    \caption{$t_{max}$ and $\Delta T$ vs power savings for the $1.863$ radians, $2.256$ radians, and daylight trajectories.}
    \label{fig:dt_vs_tmax_all}
\end{figure*}

In \Fig{fig:dt_vs_tmax_all}, we list the graphs comparing $t_{max}$, $\Delta T$, and power savings for the 1.863 radians, 2.256 radians, and daylight trajectories.
The graph for the 1.47 radian trajectory can be found in \Fig{fig:validation:dt_vs_tmax} of the main text.
All results assume a $\Delta T$ of 5 JND.

\section{3 JND Results}
\label{sec:3jnd}

\begin{figure}[h]
    \centering
    \includegraphics[width=0.35\linewidth]{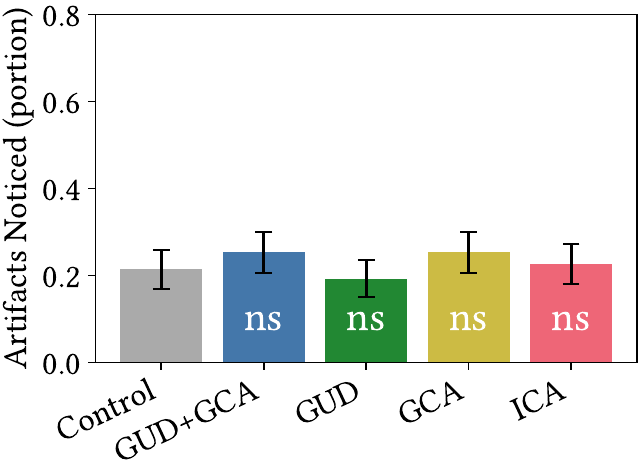}
    \caption{A comparison of participants' ability to notice artifact in each condition. Here, we benchmark 20\% power savings, which corresponds to a $\Delta T$ allowance of 3 JND. No algorithms are noticed statistically more than control.}
    \label{fig:3jnd_results}
\end{figure}

We benchmark the difference between the different algorithms with a 20\% power savings target, which corresponds to a $\Delta T$ allowance of 3 JND.
The results are displayed in \Fig{fig:3jnd_results}.
Brightness Rolloff (BR) is excluded from these tests due to its poor performance in the prior round.
None of the algorithms benchmarked are noticed statistically more frequently than control ($p > 0.05$). 
\section{ROC Curves}
\label{sec:rocs}

\begin{figure}[h]
    \centering%
    \subfloat[GUD+GCA, 5 JND (31\% savings)]{
        \includegraphics[height=1.8in]{figures/roc_Both5-120.pdf}
        \label{fig:rocs:both5}
    }
    \subfloat[GUD, 5 JND (31\% savings)]{
        \includegraphics[height=1.8in]{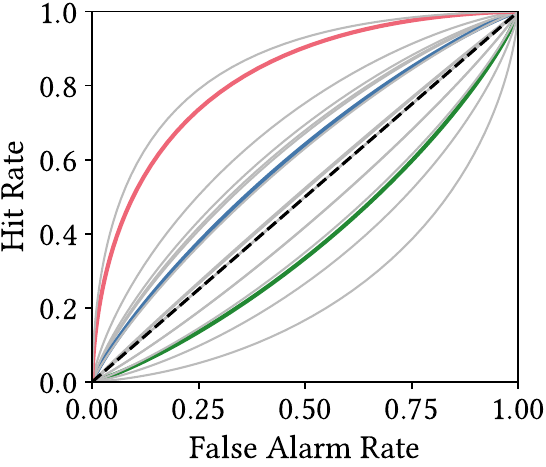}
        \label{fig:rocs:gud5}
    }%
    \subfloat[GCA, 5 JND (31\% savings)]{
        \includegraphics[height=1.8in]{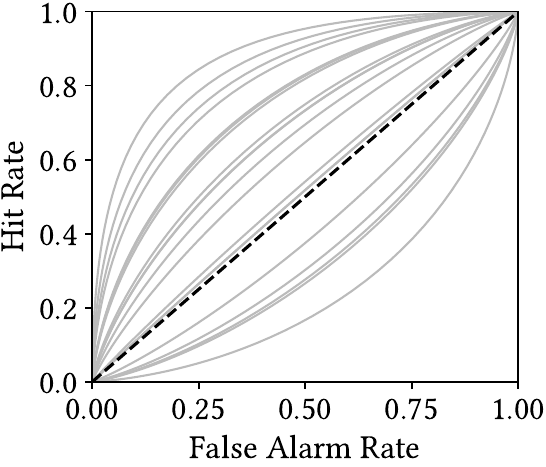}
    }\\
    \vspace{5pt}
    \centering%
    \subfloat[BR, 5 JND (31\% savings)]{
        \includegraphics[height=1.8in]{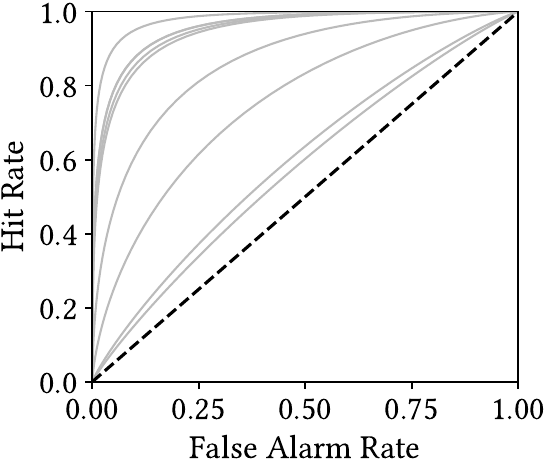}
    }%
    \subfloat[ICA, 5 JND (31\% savings)]{
        \includegraphics[height=1.8in]{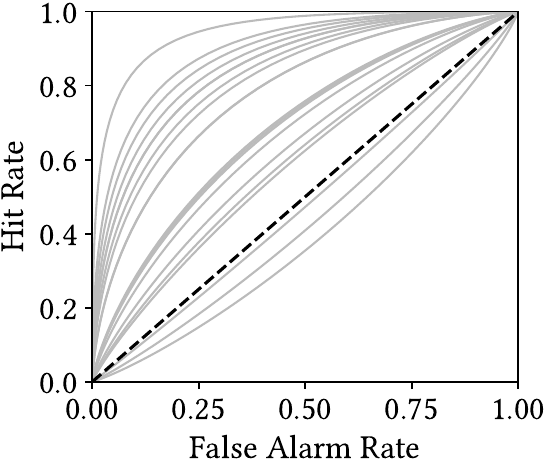}
    }%
    \subfloat[GUD+GCA, 3 JND (20\% savings)] {
        \includegraphics[height=1.8in]{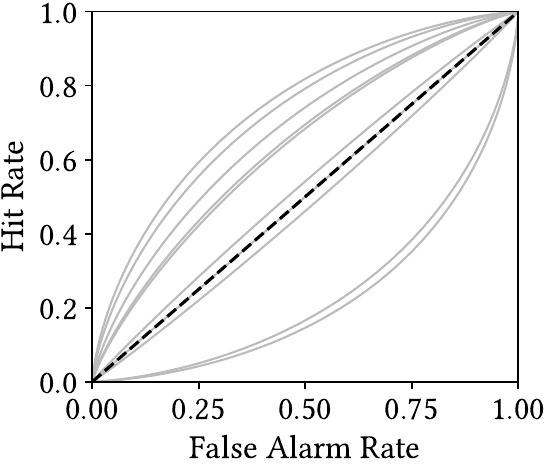}
    }\\
    \vspace{5pt}
    \centering%
    \subfloat[GUD, 3 JND (20\% savings)] {
        \includegraphics[height=1.8in]{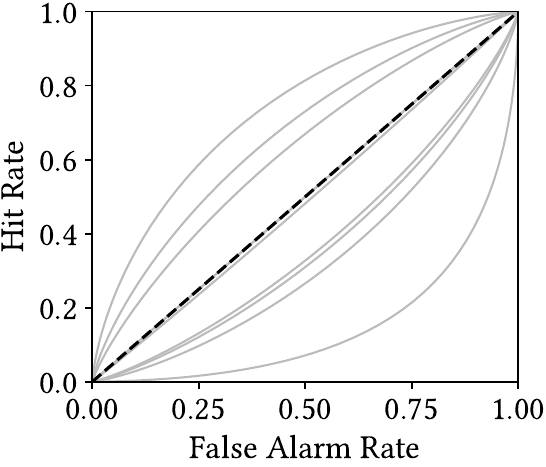}
    }%
    \subfloat[GCA, 3 JND (20\% savings)] {
        \includegraphics[height=1.8in]{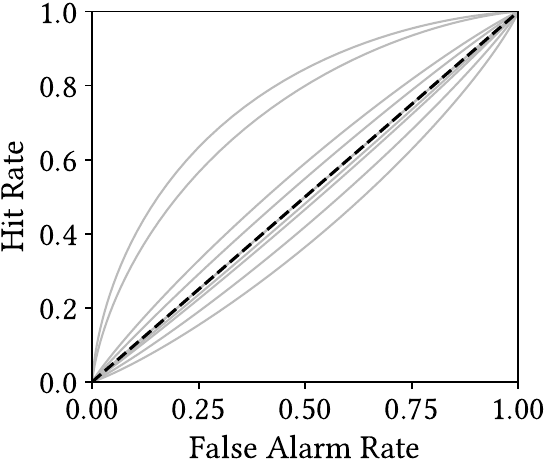}
    }%
    \subfloat[ICA, 3 JND (20\% savings)] {
        \includegraphics[height=1.8in]{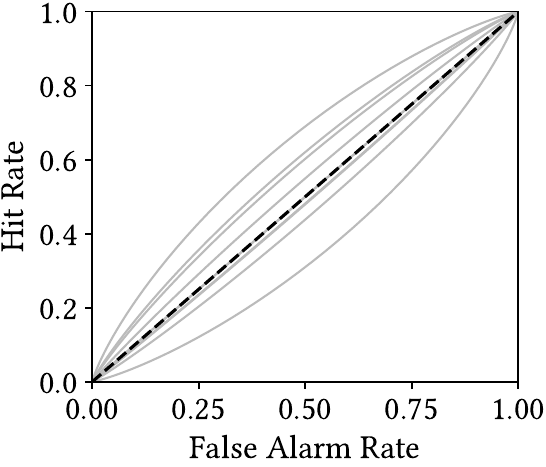}
    }%
    \caption{ROC curves for 5 JND (31\% power savings) and 3 JND (20\% power savings) threshold for $\Delta T$.}
    \label{fig:rocs:rocs}
\end{figure}

The complete set of Receiver Operating Characteristic (ROC) curves are graphed in \Fig{fig:rocs:rocs}.
Three participants that were able to spot the GUD+GCA solution exceptionally well are highlighted in \Fig{fig:rocs:both5}.
These same participants are highlighted again in the ROC curve for GUD, \Fig{fig:rocs:gud5}.
Two of the three participants are not able to detect GUD well.

\newpage
\section{$d'$ Bar Charts}
\label{sec:dp}

\begin{figure}[h]
    \centering%
    \subfloat[d' values for 5 JND (31\% savings)] {
        \includegraphics[width=3.364431853in]{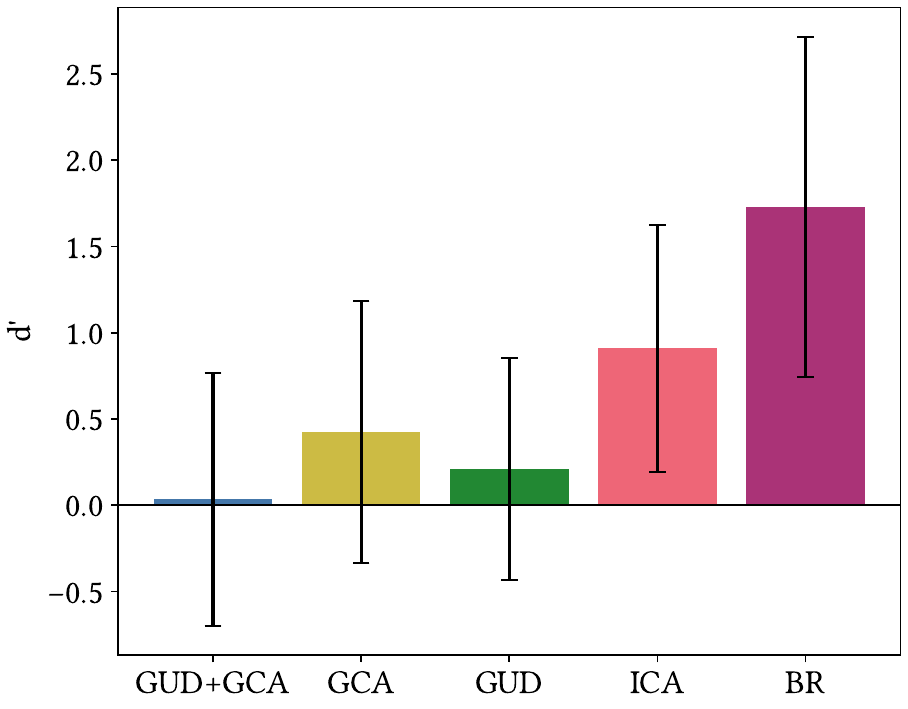}
    }%
    \subfloat[d' values for 3 JND (20\% savings)] {
        \includegraphics[width=3.364431853in]{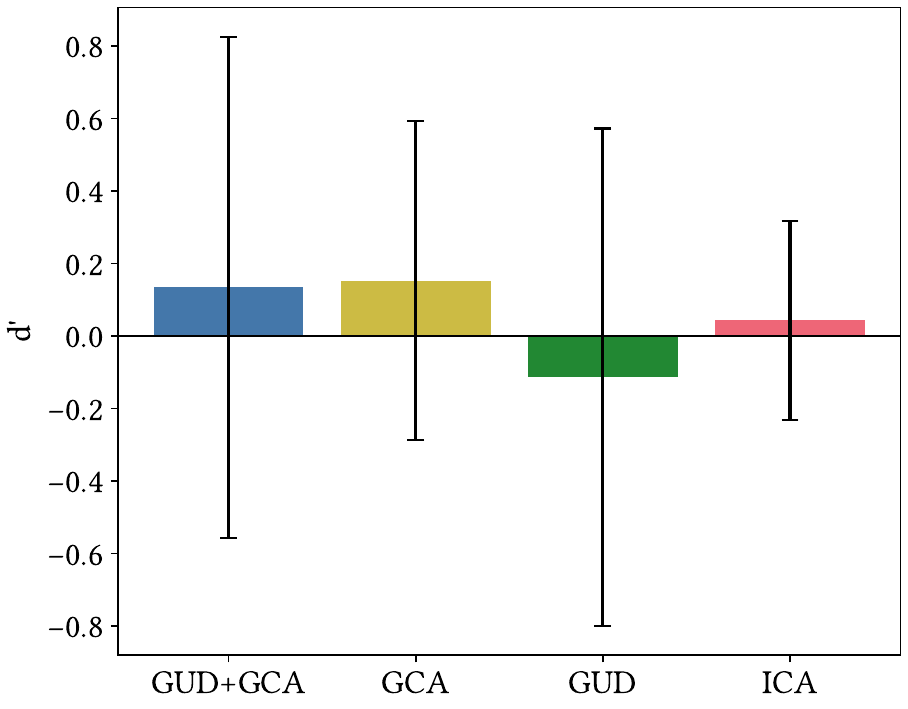}
    }%
    \caption{$d'$ values for 31\% and 20\% power savings. Error bars indicate one standard deviation.}
    \label{fig:dp:dp}
\end{figure}
\section{$k_1$ $k_2$ Fit Uncertainty}
\label{sec:uncertainty}

\begin{figure}[h]
    \centering%
    \subfloat[\fixedme{$k_1$ confidence interval}] {
        \includegraphics[width=3.364431853in]{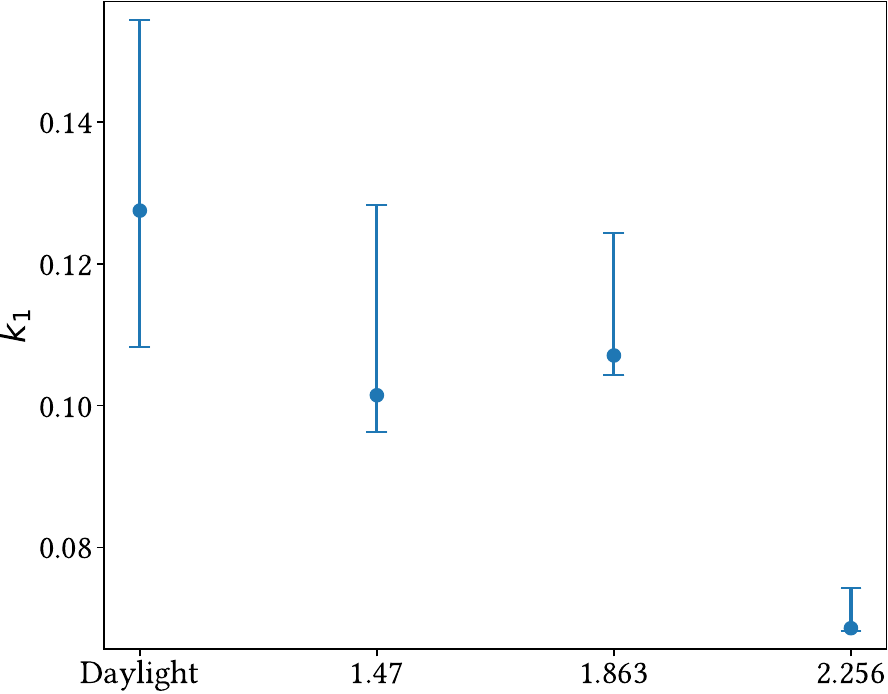}
        \label{fig:confidence:k1}
    }%
    \subfloat[\fixedme{$k_2$ confidence interval}] {
        \includegraphics[width=3.364431853in]{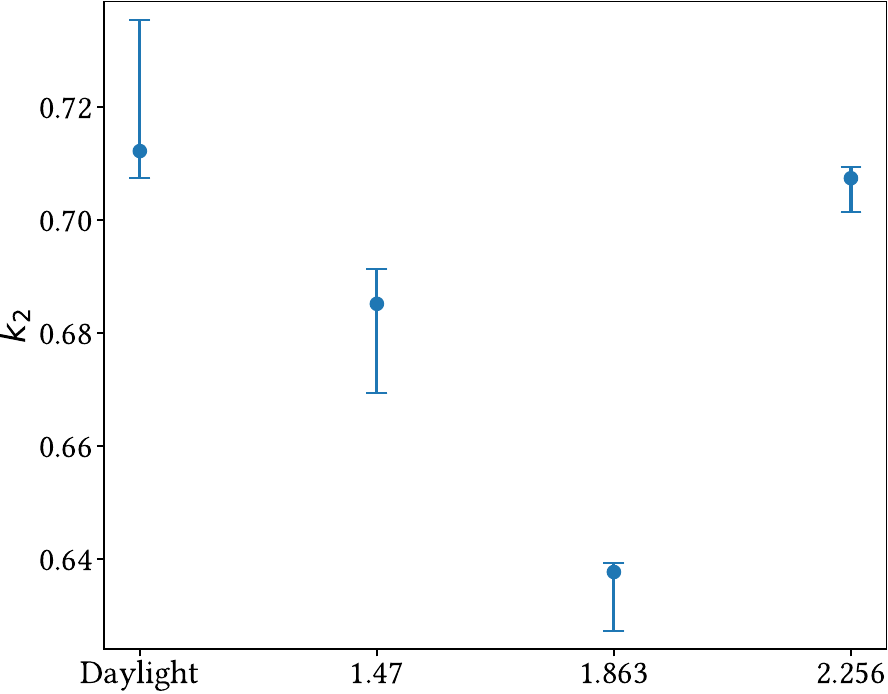}
        \label{fig:confidence:k2}
    }%
    \caption{\fixedme{$k_1$ and $k_2$ $95\%$ confidence intervals. The upper and lower whiskers of each plot represent the upper and lower bounds of the $95\%$ confidence interval. The dot represents the value at which the MLE equation is evaluated to be highest.}}
\end{figure}

\fixedme{
    The $95\%$ confidence intervals for the $k_1$ and $k_2$ fits in \Tbl{pilot_results:k1k2} are graphed in \Fig{fig:confidence:k1} and \Fig{fig:confidence:k2}.
    For each trajectory, we lump the data across all participants and all velocities and fit the corresponding $k_1$ and $k_2$ coefficient.
}

\fixedme{
Note that the $k_1$ and $k_2$ figures above are fit on a population-level, not on an individual-by-individual basis.
While the $95\%$ confidence intervals are small for the population-level fit, we do not have confident fits for individuals.
There is not a sufficient amount of data per-individual to obtain a good estimate of $k_1$ and $k_2$.
The standard deviations and average confidence interval widths of \textbf{individually-fit} $k_1$ and $k_2$ values by trajectory are documented in \Tbl{tab:uncertainty:individual}.
}

\fixedme{
\begin{table*}
    \caption{Variance and mean confidence interval widths for \textbf{individual-level} fits of $k_1$ and $k_2$.}
    \centering
    \begin{tabular}{c|c|c|c|c}
        \toprule[0.10em]
        Trajectory & $k_1$ std & $k_1$ mean CI width & $k_2$ std & $k_2$ mean CI width \\ 
        \midrule[0.05em]
        Daylight & 0.256 & 0.202 & 0.0962 & 0.0653 \\
        1.470 & 0.207 & 0.340 & 0.0794 & 0.0717 \\
        1.863 & 0.382 & 0.322 & 0.115 & 0.0890 \\
        2.256 & 0.369 & 0.273 & 0.109 & 0.0842 \\
        \bottomrule[0.10em]
    \end{tabular}
    \label{tab:uncertainty:individual}
\end{table*}
}
\section{\fixedme{Accounting for dynamic lighting}}
\label{sec:dynamic}
\fixedme{
Using a more complex problem formulation, we are able to account for VR scenes containing dynamic lighting as well. 
Let $A_o(t)$ represent the illuminant over time of the origninal VR content.
Let $a_o(t)$ represent the predicted adaptation state of a user that is exposed to the original VR content.
In our formulation, $A(t)$ is the target illuminant of the modified VR scene, which we hope to solve for.
$a(t)$ remains defined as the predicted adaptation state of a user exposed to VR content modified by out algorithm.
}

\fixedme{
We want to constrain $A(t)$ such that the appearance of the illuminant $A(t)$ under the adaptation state $a(t)$ is similar to the appearance of $A_o(t)$ under $a_o(t)$.
Note that, since the lighting of the background in the original VR content is now dynamic, the user's perception of white may no longer align with the illuminant of the scene, even in the original content. 
This constraint translates to the following formulae:
}

\fixedme{
\begin{align}
    A^*(t) = f_{a_o(t) \to a(t)}(A_o(t)) \\
    \left|A(t) - A^*(t)\right| \leq \Delta D
\end{align}
}

\fixedme{
In the above formula, $A^*(t)$ represents the color that appears like the illuminant $A_o(t)$ under the adaptation state $a_o(t)$, to an observer with adaptation state $a(t)$. $f_{a_o(t) \to a(t)}$ represents the Chromatic Adaptation Transform (CAT) function. $\Delta D$ is similar in definition to $\Delta T$ -- it is the maximum permissible difference between the rendered illuminant, $A(t)$, and the perceptually accurate illuminant, $A^*(t)$.
}

\fixedme{
Hence, the new optimization formulation (in lieu of \Equ{eq:optimization:opt_func} in the main text) is:
}

\fixedme{
\begin{align}
    \arg \min_{A(t)} \text{Power}(A(t)) \\
    \text{s.t.} \forall t, \left| A(t) - A^*(t) \right| \leq \Delta D
\end{align}
}

\end{document}